\documentclass[a4paper,12pt]{article}
\usepackage[english]{babel}
\usepackage[ansinew]{inputenc}
\usepackage{eurosym}
\usepackage{amsfonts}
\usepackage{mathrsfs}
\usepackage{amsthm}
\usepackage{amsmath, amssymb, bm}
\usepackage{mathtools}
\usepackage{graphics}
\graphicspath{{graphics/}}
\usepackage[usenames,dvipsnames]{xcolor}
\usepackage{epsfig}
\usepackage{rotating}
\usepackage{threeparttable}
\usepackage{booktabs}
\usepackage{multirow}
\usepackage{dcolumn}
\usepackage{hyperref}
\usepackage{apacite}
\usepackage[width=15cm,labelfont=bf]{caption}
\usepackage{pgfplots}
\usepackage{diagbox}

\pgfplotsset{width=10cm,compat=1.9}
\usepackage{calrsfs}
\DeclareMathAlphabet{\pazocal}{OMS}{zplm}{m}{n}

    \def\0{\mbox{\bf{0}}}

\usepackage{graphicx}
\usepackage{apacite}
\usepackage{verbatim}
\usepackage{setspace}
\usepackage{caption}
\usepackage{subcaption}
\usepackage{longtable}
\usepackage{colortbl}
\usepackage{xtab}
\usepackage{tikz,booktabs}
\usepackage[bottom]{footmisc}
\usepackage{indentfirst}
\usepackage{endnotes}
\usepackage{rotating}
\usepackage{lscape}
\usepackage{color, colortbl}%
\usepackage{authblk}
\usepackage{mdwlist}

\usepackage{algpseudocode}
\usepackage{color, colortbl}%
\usepackage{authblk}
\usepackage{algorithm}
\usepackage{algorithmicx}
\definecolor{Gray}{gray}{0.9}

\begin{document}
	
	\title{\centering The Forecasting Performance of  Factor Models with Martingale Difference
		errors\thanks{The authors are grateful to Tommaso Proietti and his valuable and insightful comments, which led to several improvements in both the presentation and the content of the paper. The authors are also grateful to the participants of the 41st International Symposium on Forecasting 2022. Alessandro Giovannelli gratefully acknowledges financial support from the Italian Ministry of Education, University and Research, Progetti di Ricerca di Interesse Nazionale, research project 2020-2023, project 2020N9YFFE. All numerical simulations in this article have been realized on the Linux HPC cluster Caliban located in the Laboratory of High-Performance Parallel Computing at the University of L'Aquila.} }

 \author[1]{L. M. Rolla}
	\author[2]{A. Giovannelli}

\affil[1]{\small University of Rome ``Tor Vergata", Italy}
\affil[2]{\small University of L'Aquila, Italy}

	
	
	\date{}
	\maketitle
	
	\begin{abstract}
		\noindent  This paper analyses the forecasting performance of a new class of factor models with martingale difference errors (FMMDE) recently introduced by \cite{lee2018martingale}. The FMMDE makes it possible to retrieve a transformation of the original series so that the resulting variables can be partitioned according to whether they are conditionally mean-independent with respect to past information. We contribute to the literature in two respects. First, we propose a novel methodology for selecting the number of factors in FMMDE. Through simulation experiments, we show the good performance of our approach for finite samples for various panel data specifications. Second, we compare the forecasting performance of FMMDE with alternative factor model specifications by conducting an extensive forecasting exercise using FRED-MD, a comprehensive monthly macroeconomic database for the US economy. Our empirical findings indicate that FMMDE provides an advantage in predicting the evolution of the real sector of the economy when the novel methodology for factor selection is adopted. These results are confirmed for key aggregates such as Production and Income, the Labor Market, and Consumption.

	\end{abstract}
	
	
	\newpage
	
	\section{Introduction}
	
	\noindent This paper investigates the forecasting performance of factor models with martingale difference error (FMMDE, \citeA{lee2018martingale}), a recent advancement in factor modeling literature. The main features of FMMDE are outlined below. Under the assumption that the idiosyncratic component is a martingale difference sequence - a more stringent condition than just assuming that it follows a white noise process - it is possible to show that there exists a contemporaneous linear transformation of the original series such that the resulting series are segmented into two groups, one of being conditionally mean independent upon past information. This property of the model has important implications in terms of optimal prediction, as it reduces the modeling effort to the subset of factors that exhibit some form of serial dependence in the conditional mean. Moreover, employing FMMDE is particularly advantageous when analyzing time series that exhibit nonlinear dependencies, as the model relies upon measures of statistical dependence that are suitable for detecting nonlinear relationships in the data. In this regard, we provide simulated examples that highlight the good performance of FMMDE in terms of forecast accuracy within nonlinear contexts.

 \vspace{0.3cm}
 
\noindent We contribute to the existing literature in two respects. First, we introduce a novel technique for determining the dimension of the latent factor process. Our method, in particular, rests upon a sequential testing procedure to identify those transformed series for which the martingale difference hypothesis can be rejected, i.e., those exhibiting some form of conditional mean dependence upon the past. Through an extensive Monte Carlo simulation exercise, we show an improved accuracy in the estimation of the true number of factors as we compare our method to the eigenvalue ratio procedure outlined in Lee and Shao's original paper. The evidence is confirmed in high-dimensional settings when the number of variables greatly exceeds the number of observations.
 
\noindent  Second, we assess the forecasting performance of FMMDE to establish whether it can represent a valid alternative when predicting macroeconomic aggregates. For this purpose, we compare its empirical properties to those of the factor model defined in Stock and Watson ( SW hereafter), see \citeA{stock2002forecasting}, \citeA{stock2002macroeconomic}. Furthermore, we provide a comparison with the model proposed by \cite{lam2011estimation}, hereafter referred to as LYB, which represents an analog of FMMDE in the case where the idiosyncratic term is white noise by construction, and the factors are assumed to exhibit serial linear dependence. The two models, FMMDE and LYB, share a significant number of features, as they both rely on the information obtained from specific autocovariance matrices at non-zero lags. For this class of models, the factors capture all the non-trivial dynamics of the data, while cross-sectional dependence may be explained by both the factors and the idiosyncratic components. This approach differs from the one followed by more traditional techniques, which rely upon the information contained in the variance-covariance matrix. In this case, in fact, the factors capture most of the cross-sectional dependence, while the idiosyncratic terms, not subjected to serial independence constraints, may contain some non-trivial dynamics.

 \noindent The estimation of FMMDE rests upon the definition of martingale difference divergence matrix, which can be regarded as a generalization of the standard autocovariance matrix based upon measures of nonlinear serial dependence. In particular, the martingale difference divergence matrix encodes the information about the number and form of linear combinations of the original vector of time series that are conditional mean independent of past information. We can retrieve this set of transformations by performing principal component analysis on the sample martingale difference divergence matrix. The estimation involves selecting the value of a parameter to determine the lag at which we truncate the serial information used by the autocovariances. Since the original paper provides no optimal way to determine the parameter, in our empirical analysis, we resort to a cross-validation procedure to determine the most appropriate value to be used at each time interval, following a squared forecast error minimization criterion.

 \vspace{0.3cm}

\noindent  Using a comprehensive US macroeconomic dataset, the FRED-MD \footnote{The dataset, introduced by \cite{mccracken2016fred}, consists of 123 monthly macroeconomic time series observed from January 1959 to December 2019. We remind to the enclosed Appendix for more information about the nature and composition of the series in the dataset.}, we assess the forecasting performance of the aforementioned techniques by conducting a pseudo-real-time exercise with forecast horizon at 1, 6, and 12 months ahead. In addition to the usual target variables such as Industrial Production, Unemployment Rate, All Employees, and Consumer Price Index, the prediction exercise is also conducted on the remaining variables composing the dataset. In this way, it is possible to get a more general picture of the different sectors beyond the forecasting performance on the individual macroeconomic aggregates.

 \noindent Comparing results obtained for FMMDE using the sequential testing and eigenvalue ratio procedures for factor selection, our empirical evidence suggests that the former performs better for the Output and Labour Market sectors. On the other hand, the eigenvalue ratio procedure leads to better results in predicting the Prices sector, especially for the medium to long horizons. We attribute this difference in the results to the generally more considerable number of factors retrieved by the sequential testing method, which gives an advantage in predicting real economic activity measures. The previous results are confirmed when looking at the forecast performance of key macroeconomic indicators such as Industrial Production, Unemployment Rate, and Consumer Price Index. Comparing the performance of FMMDE with the factor models introduced by SW, it emerges that the two generally exhibit similar performance for short-term forecasting horizons. However, for the medium to long horizons, using FMMDE with the sequential testing procedure offers significant advantages. 
	\vspace{0.3cm}
 
	\noindent  The paper is structured as follows. We introduce the main theoretical properties of
	the FMMDE and the nonlinear dependence measures used in its definition in Section \ref{section:FMMDEmodel}. In particular, we discuss the martingale difference divergence matrix, clarifying its role as a generalized covariance measure. Section \ref{sec:factorSelection} details the sequential testing methodology employed to estimate the number of factors in the model. We compare the finite sample performance of our method to that of the original eigenvalue ratio methodology through simulation examples. Section \ref{section:SWmodel} briefly discusses the SW model and its main properties. Additionally, it provides an introduction to the LYB model, detailing its close relationship with FMMDE. We study the models' predictive capabilities in both linear and nonlinear settings. The composition of the dataset, estimation methodology, and forecasting procedure are treated in Section \ref{section:EmpAna}. In the same section, we define the cross-validation strategy used for selecting the value of the lag parameter used by the martingale difference divergence matrix in the estimation of FMMDE. Finally, the forecasting performance of the competing models is discussed in Section \ref{section:empirical results}, where we report results obtained for both individual series and the entire dataset as an aggregate.

	\section{Factor Model with Martingale Difference Error}
	
	\label{section:FMMDEmodel}
	
	\noindent Let $\mathbf{x}_{t}=(x_{1t},x_{2t},...,x_{nt})^{\prime}\in\mathbb{R}^{n}$ be a strictly stationary time series, and $%
	\mathcal{F}%
	_{t-1}=\sigma\left(  \mathbf{x}_{t-1},...,\mathbf{x}_{1},...\right)  $
	be the past information set. It is well known that the best predictor of $\mathbf{x}_t$, in the mean squared error sense, coincides with the conditional mean $E\left(\mathbf{x}_t| \mathcal{F}_{t-1} \right) $.
 In the original paper, \cite{lee2018martingale} demonstrate how it is possible to identify a contemporaneous
	linear transformation, denoted as $\mathbf{M}%
	$\textbf{ }$\in$ $R^{n\times n}$, such that $\mathbf{Mx}_{t}$ $=$ $\left[
	\mathbf{F}_{a,t}^{\prime}\mathbf{,F}_{b,t}^{\prime}\right]  ^{\prime}$, where
	$E\left(  \mathbf{F}_{a,t}|%
	\mathcal{F}%
	_{t-1}\right)  \neq E\left(  \mathbf{F}_{a,t}\right)  $ and $E\left(
	\mathbf{F}_{b,t}|%
	\mathcal{F}%
	_{t-1}\right)  $ $=E\left(  \mathbf{F}_{b,t}\right)  $ with $\mathbf{F}_{a,t}%
	$\textbf{ }$\in$ $\mathbb{R}^{r}$ and $\mathbf{F}_{b,t}\in\mathbb{R}^{n-r}$.
	In other words, the resulting series
	can be separated into two parts, with one part being
	conditionally mean dependent upon $%
	\mathcal{F}%
	_{t-1}$, and the other being conditionally mean
	independent upon the past information set. The modeling task
	for the entire series $\mathbf{x}_{t}$ will be therefore reduced to that for the lower dimensional
	series $\mathbf{F}_{a,t}$ without any loss in terms of predictive accuracy.

	\noindent Such a problem can be formulated equivalently
	in a factor model framework: under the assumption that
	the variables have a latent factor structure; the approximate factor model is
	defined as:
	\begin{equation}
		\mathbf{x}_{t}=\mathbf{\Lambda F}_{t}+\boldsymbol{\zeta}\mathbf{_{t}}\text{,}%
		\label{eq:model}
	\end{equation}
	where $\mathbf{F}_{t}=(F_{1t},...,F_{rt})'$ are $r<n$
	unobserved common factors, $\mathbf{\Lambda}\in\mathbb{R}^{n \times r}$ is the matrix of
	factor loadings and $\boldsymbol{\zeta}_{t}=(\zeta_{1t},...,\zeta
	_{nt})^{\prime}\in\mathbb{R}^{n}$ represents the idiosyncratic disturbance not explained by the
	factors. The factor model with martingale difference error can be
	expressed as%
	\begin{equation}
		\mathbf{x}_{t}=E\left(  \mathbf{x}_{t}|%
		\mathcal{F}%
		_{t-1}\right)  +\boldsymbol{\zeta}_{t}\text{,} \label{iewfg}%
	\end{equation}
	with $E(\mathbf{x}_{t}|%
	\mathcal{F}%
	_{t-1})=\mathbf{\Lambda}\mathbf{F}_{t}$, where, by construction, the idiosyncratic component $\boldsymbol{\zeta
	}_{t}$ is a vector martingale
	difference sequence. Defining $\mathbf{M}$ as the partitioned matrix $\mathbf{M}=\left[ \mathbf{\Lambda 
	},\mathcal{B}\right] ^{\prime }$, for $\mathcal{B} \in \mathbb{R}^{n\times \left( n-r\right) }$ being a matrix for
	which the relationship $\mathbf{\Lambda }^{\prime }\mathcal{B=}$ $0$ holds
	true, we can retrieve
	an estimate of the factor process as a result of obtaining an estimate of the matrix $\widehat{\mathbf{M}}$  as the transformation $\mathbf{\widehat{\mathbf{F}}}_{t}=%
	\widehat{\mathbf{\Lambda}}^{\prime}\mathbf{x}_{t}.$ The identification of $\mathbf{M}$ is
	made feasible by the introduction of martingale difference divergence
	matrix, which allows us to express $\mathbf{x}_{t}$ as
	a nonsingular linear transformation of a $r$-dimensional dynamically
	dependent common factor process, and a martingale difference error component
	of dimension $(n-r)$. Given matrix can be regarded as a multivariate extension of the notion of
	martingale difference divergence, introduced in \cite{shao2014martingale} as a measure of conditional mean independence between random vectors and defined as (the non-negative square root of)
	\begin{equation}
		MDD\left(  {x}_{h,t}|\mathbf{x}_{t-j}\right)^{2}  =\frac{1}{c_{n}}%
		\int_{\mathbb{R}^{n}}\frac{\left\vert G_{h}\left(  \mathbf{s}\right) \right\vert^{2} }{\left\vert \mathbf{s}\right\vert _{n}^{1+n}}d\mathbf{s}
	\end{equation}
	where $G_{h}\left(
	\mathbf{s}\right)  =$ cov$\left(  x_{h,t},\exp i\left\langle \mathbf{s},\mathbf{x}_{t-j}%
	\right\rangle \right)  $    and $c_{n}=\pi ^{\left( 1+n\right) /2}/\Gamma \left( \left( 1+n\right) /2\right)   
	$.\footnote{Here we denote
		$i=\sqrt{-1}$. For a complex-valued function $f(.)$ the complex conjugate of $f$ is
		denoted by $f^{\ast}$ and $|f|^{2}=ff^{\ast}.$} \footnote{Martingale difference divergence can be regarded as an extension of the concept distance covariance (\cite{szekely2007measuring}) using a similar weighting function under the integral.} In our case, martingale difference divergence can be used to measure the departure from the conditional mean independence relationship 
	$E\left( x_{h,t}|\mathbf{x}_{t-j}\right) =E\left( x_{h,t}\right)$ almost surely, for a fixed
	$j\in\mathbb{N}^{+}$, and for each $h$-th element in the vector $\mathbf{x}_{t}$. In particular, martingale difference divergence is defined so that conditional mean independence is verified if and only if $	MDD\left(  {x}_{h,t}|\mathbf{x}_{t-j}\right)^{2}  =0$. The function $G_{h}\left( \cdot \right)  $ can be interpreted as a generalized covariance function between $h$-th element in $\mathbf{x}_{t}$ and the complex exponential of its lag: this allows to express dependence conditions in terms of characteristic functions of the variable in question. As discussed in \cite{hong1999hypothesis}, characteristic functions can be used to check for a variety of dependence conditions different from simple correlation, allowing, in particular, to assess the existence of nonlinear dependence relations. It is well known that, in the case of non-Gaussian and nonlinear time series, the autocorrelation function cannot fully characterize dependence structures. In this sense, MDD allows us to assess the existence of conditional mean dependence in the absence of linear relationships when only nonlinear dependence is present.

	\noindent In the case of
	multivariate time series objects, the martingale difference divergence matrix is defined as the $n\times n$ matrix%
	\begin{equation}
		\mathbf{MDDM}\left(  \mathbf{x}_{t}|\mathbf{x}_{t-j}\right)  =\frac{1}{c_{n}}%
		\int_{\mathbb{R}^{n}}\frac{\mathbf{G}\left(  \mathbf{s}\right)  \mathbf{G}\left(  \mathbf{s}\right)  ^{\ast}%
		}{\left\vert \mathbf{s}\right\vert _{n}^{1+n}}d\mathbf{s}\text{,}%
	\end{equation}
	where $\mathbf{G}\left(  \mathbf{s}\right)  =$ cov$\left(  \mathbf{x}_{t},\exp
	i\left\langle \mathbf{s},\mathbf{x}_{t-j}\right\rangle \right)  =\left(  G_{1}\left(
	\mathbf{s}\right)  ,...,G_{n}\left(  \mathbf{s}\right)  \right)  ^{\prime}$ for $\mathbf{s}\in
	\mathbb{R}^{n}$. For $h=1,...,n$, elementwise, we have\ $G_{h}\left(
	\mathbf{s}\right)  =$ cov$\left(  x_{ht},\exp i\left\langle \mathbf{s},\mathbf{x}_{t-j}%
	\right\rangle \right)  $. Accordingly, its $(i,i)$th entry  equals to
	$MDD(x_{it}|\mathbf{x}_{t-j})^{2}$.

	\noindent Assuming that $E\left(  |\mathbf{x}_{t}|_{n}%
	^{2}+|\mathbf{x}_{t-j}|_{n}^{2}\right)  $ $<\infty$, the matrix is real, symmetric, and positive semi-definite, so that, for any real matrix $\mathbf{\Lambda}\in
	\mathbb{R}^{n\times r}$ we observe the relationship $	\mathbf{MDDM}\left(  \mathbf{\Lambda}^{\prime}\mathbf{x}_{t}|\mathbf{x}_{t-j}\right)
	=\mathbf{\Lambda}^{\prime}\mathbf{MDDM}(\mathbf{x}_{t}|\mathbf{x}_{t-j})\mathbf{\Lambda
	}$. There exist, subsequently, $\left(n-r\right)$ linearly independent combinations that are
	conditionally mean independent of $\mathbf{x}_{t-j}$, i.e., $MDD\left(
	\mathcal{B}_{i}^{\prime}\mathbf{x}_{t}|\mathbf{x}_{t-j}\right)  =0$, $i=1,...,n-r$, if and only if Rank$\left( \mathbf{MDDM}\left(  \mathbf{x}%
	_{t}|\mathbf{x}_{t-j}\right)  \right)  =r$. In this sense, the martingale difference divergence matrix can be employed to determine the set of linear combinations that are conditional mean
	independent of past information. Let $\left\{  \lambda_{i},\gamma_{i}\right\}  _{i=1}^{n}$ \ be the $n$ pairs of
	eigenvalues and eigenvectors of $\mathbf{\Gamma}_{k_{0}}$, the matrix
	$\mathbf{M}=\left[  \mathbf{\Lambda},\mathcal{B}\right]  ^{\prime}$ can be
	obtained 
	by the spectral analysis of $\mathbf{\Gamma}_{k_{0}}$ as the matrix of its eigenvectors, so that $\mathbf{M=}\left[ \gamma _{1},...,\gamma _{n}\right] $. In particular,
	$\mathbf{\Lambda}$ will be the matrix collecting eigenvectors corresponding to
	the $r$ factors that are not martingale difference sequences; conversely, $\mathcal{B}$ will correspond to the
	remaining $n-r$ eigenvectors for which $MDD\left(  \mathcal{B}_{i}^{\prime
	}\mathbf{x}_{t}|\mathbf{x}_{t-j}\right)  =0$, for every
	$i=r+1,...,n,$ and for any $j \in \mathbb{N}^{+}$.

\vspace{0.3cm}
 
\noindent	For a given sequence of observations $\{\mathbf{x}_t \}^{T}_{t=1}$, defining $\mathbf{z}_{t}=\mathbf{x}_{t-j}$, for $j=1,...,T-1$, sample
	martingale difference divergence matrix can be expressed as%
	\begin{equation}
		\mathbf{MDDM}_{T}\left(  \mathbf{x}_{t}|\mathbf{x}_{t-j}\right)  =-\frac{1}{\left(T-j\right)^{2}}%
		\sum_{h,l=1}^{T-j}\left(  \mathbf{x}_{h}-\overline{\mathbf{x}}\right)  \left(
		\mathbf{x}_{l}-\overline{\mathbf{x}}\right)  ^{^{\prime}}\left\vert
		\mathbf{z}_{h}-\mathbf{z}_{l}\right\vert _{n}\text{.}%
	\end{equation} From a practical standpoint, as we have only a limited number of observations available, we need to approximate the past information set,$\ \mathcal{F}_{t-1}%
	$, with its finite sample
	analogue $\mathcal{F}_{t-1,t-k_{0}}=\sigma\left(  \mathbf{x}_{t-1}%
	,...\mathbf{x}_{t-k_{0}}\right)$, with $k_{0}$ being a pre-specified integer. For this purpose, the original paper introduced the concept of cumulative martingale difference divergence matrix to
	quantify the conditional mean independence of $\mathbf{x}_{t}$ on its recent
	past $\mathcal{F}_{t-1,t-k_{0}}$, defined as%
	\begin{equation}
		\mathbf{\widehat{\Gamma}}_{k_{0}}=\sum_{j=1}^{k_{0}}\mathbf{MDDM}_{T}\left(  \mathbf{x}%
		_{t}|\mathbf{x}_{t-j}\right)  \text{.}%
		\label{eq:CMDDM}
	\end{equation} This reduces the analysis to that of the matrix
	$\widehat{\mathbf{M}}$ $=\left[  \widehat{\mathbf{\Lambda}},\mathcal{\widehat
		{\mathcal{B}}}\right]  ^{\prime}$ of the eigenvectors of
	$\mathbf{\widehat{\Gamma}}_{k_{0}}$. From now on, we mainly refer to the vector of transformed variables $\widehat{\mathbf{P}}_t =  \widehat{\mathbf{M}} \mathbf{x}_t$. By construction, such linear transformation gives rise to a separation between the estimated factors and a number of MDS components, so that
	$\widehat{\mathbf{P}}_t =  \left[  \widehat{\mathbf{F}}_t^{\prime},\widehat
	{\mathbf{E}}_t^{\prime} \right]  ^{\prime}\in$ $\mathbb{R}^{n}$, with $\widehat{\mathbf{F}}_{t}%
	$\textbf{ }$\in$ $\mathbb{R}^{r}$ and $\widehat{\mathbf{E}}_{t}\in\mathbb{R}^{n-r}$.

	\section{Determining the number of factors}
	\label{sec:factorSelection}
	\noindent The methodology for factor selection adopted in the original
	paper is based upon the eigenvalue ratio estimator defined in \cite{lam2011estimation}, and \cite{lam2012factor}. Let $\left\{  \widehat{\lambda}%
	_{i},\widehat{\gamma}_{i}\right\}  _{i=1}^{n}$ be the $n$ pairs of
	eigenvalues and eigenvectors obtained by the spectral analysis of
	$\mathbf{\widehat{\Gamma}}_{k_{0}}$ rearranged in order of decreasing eigenvalues. It is possible estimate $r$, i.e., the rank of
	$\mathbf{\Gamma}_{k_{0}}$, using the ratio based estimator %
	\begin{equation}
		\widehat{r}=\left(  \arg\min_{1\leq i\leq R}\frac{\widehat{\lambda}_{i+1}%
		}{\widehat{\lambda}_{i}}\right)  \text{.}%
		\label{eq:eigRatio}
	\end{equation} In the original paper, no formal rule is given for selecting the value of $R$. In general, such value can be defined as a fraction of the total number of variables $n$.
	
	\noindent We propose a novel 
	approach to determining the number of factors in the model. In particular, we
	exploit the nature of FMMDE, which allows obtaining separation of the
	unobservable series into different groups according to whether they are conditionally mean dependent upon past information or not. Specifically, we test the dynamic properties of the individual transformed series $\widehat{\mathbf{P}}_{t}=\left(\widehat{P}_{1,t},...,\widehat{P}_{n,t}\right)$ in a sequential scheme, aiming to identify the subset of those that are not a martingale difference sequence. In performing this operation, we start from the first few elements in $\widehat{\mathbf{P}}_t$, i.e., those associated with the largest estimated eigenvalues of $\mathbf{\widehat{\Gamma}}_{k_{0}}$. The testing sequence stops as soon as we are able to identify the first transformed series, $\widehat{P}_{i,t}$, for which the null hypothesis of conditional mean
	independence upon the past cannot be rejected. In this case we have $\widehat{r}=i-1$, with $\widehat{P}_{i-1,t}$ being the last element in $\widehat{\mathbf{F}}_t$ and $\widehat{P}_{i,t}$ being the first element in $\widehat{\mathbf{E}}_t$. If all transformed series are a martingale difference sequence, the testing process stops at the first iteration, $\widehat{r} = 0$, and $\widehat{\mathbf{F}}_t=\mathbf{0_v}$. 
	
	\noindent A similar approach to determining the number of factors is discussed in \cite{gao2021modeling}, which considers a generalization of the model of \cite{lam2011estimation}. In particular, estimation of the loading matrix associated with the common factors is based upon the eigenanalysis of certain autocovariance matrices, which, similarly to what happens with MDDM, allow for obtaining a separation between a set of serially dependent factors (defined in this case by a non-null degree of autocorrelation) and a white noise idiosyncratic component. The sequential testing procedure described by \cite{gao2021modeling} differs from ours for various reasons. Firstly, they investigate an alternative null hypothesis, namely the absence of serial correlation in the idiosyncratic term. In this regard, the consideration of the martingale difference hypothesis can be viewed as more comprehensive, since it allows for the possibility of observing white noise series that are not strictly martingale difference sequences. Consequently, even in the absence of linear dependence, such a series may exhibit some form of nonlinear dependence in the conditional mean. Furthermore, unlike the test statistics they utilize, the chosen test statistic in our study does not have a known limiting null distribution. As a result, we need to employ a bootstrap procedure to estimate its distribution.

 \vspace{0.3cm}
 
\noindent 	We make use of the test statistic for the martingale difference hypothesis
	introduced by \cite{wang2022testing}, defined as 
	\begin{equation}
		\widehat{T}_{wn,i}^{F}\left(  M\right)  =T\sum_{j=1}^{M}\omega_{j}\left\vert
		\left\vert \mathbf{MDDM}_{T}\left(  \widehat{P}_{i,t}|\widehat{P}_{i,t-j}\right)  \right\vert \right\vert
		_{F}\text{,}%
		\label{eq:testEq}
	\end{equation}
	where $\omega_{j}=\left(  N-j+1\right)  /\left(  Nj^{2}\right)  $ $\ $and
	$\left\vert \left\vert \cdot\right\vert \right\vert _{F}$ indicates the
	Frobenius norm.\footnote{Among all the different test specifications proposed in the
		original paper, we chose $\widehat{T}_{wn}^{F}$ as our statistic of reference as
		it represents the best compromise in terms of size and power
		properties.} For the rest of this section, to avoid using too heavy notation, we prefer, from now on, not to index the test $\widehat{T}_{wn}^{F}$ at $i$, as it is the case in Equation \ref{eq:testEq}, even though we are testing the $i$-th element in a sequential scheme. As the asymptotic null distribution of $\widehat{T}_{wn}^{F}$ is non-pivotal, the critical values of the test statistics need to be approximated by an appropriate bootstrap methodology. Following the original paper, we adopt a fixed-design wild bootstrap procedure, see \cite{wu1986jackknife} and  \cite{liu1988bootstrap}, where, given a sequence $\left\{ w_{t}^{\ast }\right\} _{t=1}^{T}$ of iid
	auxiliary random variables, and a random sample $\left\{z_t\right\}^{T}_{t=1}$, the bootstrap sample $\left\{z^{*}_t\right\}^{T}_{t=1}$ is generated as $z^{*}_t=z_tw^{*}_t$ for each $t=1,...,T$. In our specific case, we assume $\left\{ w_{t}^{\ast }\right\} _{t=1}^{T}$ to be a sequence of iid Bernoulli variates with distribution
	$
	P\left[ w_{t}^{\ast }={\scriptstyle \left(1-\sqrt{5}\right)/2}\right]  ={\scriptstyle\left(\sqrt{5}+1\right)/2%
		\sqrt{5},}$  and $
	P\left[w_{t}^{\ast }={\scriptstyle\left(1+\sqrt{5}\right)/2}\right] ={\scriptstyle\left(\sqrt{5}-1\right)/2%
		\sqrt{5}.}  
	$ Analogously to the methodology implemented in \cite{escanciano2006goodness} and \cite{wang2022testing} for testing the presence of nonlinear dependence in the residuals of a time series model, our methodology relies upon re-estimating the factor model for each of the bootstrap replicates using the bootstrap sample. The resulting vector, $\widehat{\mathbf{P}}^{*}_{t}$, will be then used to compute the bootstrap test statistic $\widehat{T}^{*}_{wn,b}$, so that, for a total of $B$ bootstrap replicates, we obtain $\left\{\widehat{T}^{*}_{wn,b}\right\}^{B}_{b=1}$, which allows us to approximate the test distribution.

	\noindent To describe the step-by-step functioning of our testing methodology, we start by illustrating the initial operation of testing the transformed series $\widehat{P}_{1,t}$, to then consider the generic $i$-th series $\widehat{P}_{i,t}$. We denote as $\mathbf{\widehat{P}}_{i:n,t}$ the vector collecting the elements of $\mathbf{\widehat{P}}_t$ from the $i$-th to the $n$-th element. The sequential testing procedure can be summarized as follows

	\begin{enumerate}

		\item We start by testing $\widehat{P}_{1,t}$, i.e., the first element in $\mathbf{\widehat{P}}_t$. Under the null hypothesis, $\widehat{P}_{1,t}$ is a martingale difference sequence: in this case, we define $\mathbf{\widehat{E}}_t=\mathbf{\widehat{P}}_t$ with $ \mathbf{\widehat{F}}_t= \mathbf{0_{v}}$. We compute the test statistic $\widehat{T}_{wn}^{F}$ for $\widehat{P}_{1,t}$.
		
		\item For each of the bootstrap replicates, we employ a sequence of iid auxiliary variables $\left\{w^{*}_{t}\right\}^{T}_{t=1}$ to generate the sequence of vectors $\{\mathbf{\widehat{E}}_t^{*} \}^{T}_{t=1}$, defined so that $\widehat{E}^{*}_{j,t}=\widehat{E}_{j,t}w^{*}_{t}$ for every $j$-th element in $\mathbf{\widehat{E}}^{*}_t$. We use it to 
		obtain the bootstrap sample  $\{ \mathbf{x}^{*}_t \}^{T}_{t=1}$ as \begin{equation}
			\mathbf{x}_{t}^{\ast}= \mathbf{\widehat{M}} \cdot \left[  \mathbf{\widehat{F}^{ \prime}}_t%
			,\mathbf{\widehat{E}}^{\ast \prime}_t\right]^{\prime},
		\end{equation}  
		
		which is then used to re-estimate the model. In particular, $\mathbf{\widehat
			{\Gamma}}^{\ast}_{k_{0}}$ is computed from the bootstrap sample: this can be used to obtain $\mathbf{\widehat{M}^{\ast}} $ and, ultimately, $\mathbf{\widehat{P}}^{\ast}_{t} $.

		\item For each of the bootstrap replicates, we compute test statistics $\widehat{T}_{wn,b}^{\ast}$ for 
		$\widehat{P}^{\ast}_{1,t}$, where the bootstrap $p$-value is computed as $\sum_{b=1}^{B}\mathbb{I}\left(
		\widehat{T}_{wn,b}^{\ast}\geqslant \widehat{T}^{F}_{wn}\right)  /\left(B+1\right)$ .

		The testing sequence comes to a halt if we are not able to reject the null hypothesis for $\widehat{P}_{1t}$ and we select $\widehat{r}=0$, accordingly. Otherwise, we proceed with testing $\widehat{P}_{2,t}$.

		\suspend{enumerate}
		Let us now define the testing procedure for $\widehat{P}_{i,t}$, i.e., the $i$-th step in the sequential procedure:
		\resume{enumerate}	\item In the sequential testing process, we were able to reject the null hypothesis up to $\widehat{P}_{i-1,t}$. In this case, we have $\mathbf{\widehat{F}}_t=\mathbf{\widehat{P}}_{1:\left(i-1\right),t}$ and $\mathbf{\widehat{E}}_t=\mathbf{\widehat{P}}_{i:n,t}$. 
		We compute the test statistic $\widehat{T}_{wn}^{F}$ for $\widehat{P}_{i,t}$.
		
		\item We proceed analogously as in points 2 and 3 above. 
		
		The testing sequence comes to a halt if we are not able to reject the null hypothesis for $\widehat{P}_{i,t}$ and we select $\widehat{r}=i-1$, accordingly. Otherwise, we proceed with testing $\widehat{P}_{i+1,t}$.

	\end{enumerate}

	\subsection{Simulation Results}
	
	\noindent In this section, we report the results of some simulation experiments to assess the performance of the proposed factor selection methodology. Results for the testing procedure are compared to those obtained for the original eigenvalue ratio methodology, defined in Equation \ref{eq:eigRatio}. The baseline model for all simulations will be
	\begin{align}
		x_{it}=\sum_{j=1}^{r}\Lambda_{ij}F_{tj}+\sqrt{\theta}\zeta_{it},\text{
			\ \ }i=1,...,n,\text{ \ }t=1,...,T.
	\end{align} Following \cite{gao2021modeling}, we assume that the factor process is characterized by an autoregressive structure
	\begin{align}
		\mathbf{F}_{t}=\mathbf{\Pi} \cdot \mathbf{F}_{t-1}+\mathbf{\eta}_{t}%
	\end{align}
	where $\mathbf{\Pi}$ is a $r\times r$ diagonal matrix with its diagonal elements being drawn independently from $U\left(  0.5,0.9\right)  ,\mathbf{\eta}	_{t}\sim N\left(  0,I_{r}\right)  $. The entries of the factor loading matrix $\mathbf{\Lambda}$ are all drawn independently from  $N\left(  0,1\right)$. Throughout the analysis we consider three data-generating processes (DGPs) for the idiosyncratic component similar to those in \cite{bai2002determining} and \cite{alessi2010improved}.

	\begin{description}
		\item 
		
		\item[$DGP_{1}$] Homoskedastic idiosyncratic components $\zeta_{it}\sim
		N\left(  0,1\right)  $
		
		\item[$DGP_{2}$] Heteroskedastic idiosyncratic components%
		\[
		\zeta_{it}=\left\{
		\begin{array}
			[c]{c}%
			\zeta_{it}^{1}\text{ \ \ \ \ \ \ \ \ \ \ if \ }t\text{ is even}\\
			\zeta_{it}^{1}+\zeta_{it}^{2}\text{ \ \ if\ \ }t\text{ is odd}%
		\end{array}
		\right.  ,\text{ \ \ \ }\zeta_{it}^{1},\zeta_{it}^{2}\sim N\left(
		0,1\right)
		\]

		\item[$DGP_{3}$] Cross-sectional correlations among idiosyncratic components%
		\[
		\zeta_{it}=v_{it}+\sum_{j\neq0,j=-J}^{J}\beta v_{i-j,t},\text{ \ \ }%
		v_{it}\text{ }\sim N\left(  0,1\right)  ,\text{ \ }\beta=0.2,\text{
			\ }J=\max\left\{  n/20,10\right\}  .
		\]

	\end{description}
	Tables from \ref{tab:simu1n50} to \ref{tab:simu3n100} report values of the empirical probabilities computed, for each methodology, in the process of determining the correct number of factors. These probabilities are obtained considering 1000 independently simulated samples, counting the number of successful factor identification attempts. Results are displayed for different values of $\theta$, $DGPs$, and for a true number of factors ranging from 2 to 5. Regarding our sequential testing procedure, we assume the fixed truncation lag $M$ - the argument used by the statistic $\widehat{T}_{wn}^{F}\left(  M\right)$ - to be equal to 6. Empirical $p$-values are computed over 499 bootstrap replications for a significance level $\alpha=5\%$. In the case of the eigenvalue ratio methodology, the value of the parameter $R$ is set to $n/3$ in all simulations. Moreover, we set $k_{0}=1$ for simplicity: our evidence suggests that results are not affected by selecting alternative values for $k_{0}$ when considering this class of models. All results are obtained after standardizing the series. The benefits of utilizing the sequential testing methodology are  particularly evident when analyzing $DGP_3$ with cross-sectional correlations among idiosyncratic components and when considering higher values of theta. The most significant results, relative to the eigenvalue ratio methodology, are in general attained for smaller values of $T$ and $n$, as it is apparent from Table \ref{tab:simu1n50} where we display results obtained for $n=50$, $T=50$ (upper part of the table), and $T=100$ (lower part of the table). It is worth noticing that better results are also obtained for $n=100$ and $T=100$ (lower part of Table \ref{tab:simu2n100}): these values of $T$ and $n$ lie approximately in the same range as those that we select for our empirical study.

 \vspace{0.3cm}
 
 \noindent We extend our simulation study by considering the model reported in Example 6 of \cite{lee2018martingale}'s paper. The three-dimensional factor space, with $\mathbf{F}_{t}=\left(
	F_{1t},F_{2t},F_{3t}\right)  ^{\prime}$, is generated from the nonlinear process $\omega_{t}$, defined by
	\begin{align}
		w_{t}=\left\{
		\begin{array}
			[c]{cc}%
			0.5+(0.05e^{-0.01w_{t-1}^{2}} & \\
			+0.9)w_{t-1}+z_{t}, & \text{if }w_{t-1}<5\text{ , }z_{t}\sim N\left(
			0,1\right) \\
			\left(  0.9e^{-10w_{t-1}^{2}}\right)  w_{t-1}+z_{t} & \text{\ if }%
			w_{t-1}\geqslant5
		\end{array}
		\right.
		\label{eq:nonlinearEQ}
	\end{align}
	with \begin{equation}
		F_{1t}   =w_{t},
		F_{2t}=w_{t-1},
		F_{3t}=w_{t-2},  \\
	\end{equation}
	Concerning the factor loading matrix $\mathbf{\Lambda}$, the first $n/2$ elements of each
	column are iid $U\left(  -2,2\right)  $ and are kept fixed once generated and
	the other elements are set to be zero. The parameter $\theta$ is now assumed to be always equal to one, while the error term
	$\boldsymbol{\zeta}\mathbf{_{t}}$ is a random sample of $N\left(  0,0.25\mathbf{\Sigma}\right)  $ independent of $\mathbf{F}_{t}$. 
	Finally, the covariance matrix $\mathbf{\Sigma=}\left(  \sigma_{i,j}\right)  _{i,j=1}^{n}$ is generated as
	\begin{equation}
		\sigma_{i,j}=\left\{
		\begin{array}
			[c]{cc}%
			\frac{1}{2}[\left(  |i-j|+1\right)  ^{2H}-\left\vert i-j\right\vert ^{2H} & \\
			+\left(  \left\vert i-j\right\vert -1\right)  ^{2H}],\text{ } & \text{if
			}i\neq j\\
			1 & \text{\ if }i=j
		\end{array}
		\right.
		\label{sigma}
	\end{equation}
	for Hurst exponent $H=0.9$. As mentioned in the original paper, the above assumptions about the dependence structure of error term make it possible to address the large $n$ case, where $n$ can exceed the number of available observations $T$. Table \ref{tab:simEx6} displays results for values $k_0 \in \left\{1,10,25\right\}$. In particular, the parameter has a non-negligible influence on the factor selection process: increasing $k_0$ improves the performance of the sequential testing method while worsening the performance of the eigenvalue ratio at the same time. Still, at the two extremes, the methods provide comparable results, with the eigenvalue ratio performing best for $k_0=1$, and the sequential testing method for $k_0=25$.
	
	\section{Comparison with other models}
	\label{section:SWmodel} 
	\noindent In this section, we investigate the forecasting performance of the FMMDE model in a simulated environment. The exercise is based on a systematic comparison with the factor model defined in Stock and Watson, see \cite{stock2002forecasting} and \cite{stock2002macroeconomic}, which can be considered a well-established benchmark in the literature and can therefore be used as a standard of reference. Letting $\mathbf{S}=T^{-1}\sum_{t}\mathbf{x}%
	_{t}\mathbf{x}_{t}^{\prime}$ and denoting the spectral decomposition of the
	covariance matrix by
	\begin{equation}
		\mathbf{S=VD}^{2}\mathbf{V}^{\prime}\text{,}%
	\end{equation}
	where $\mathbf{V}=(v_{1},\dots,v_{n})$ is the $(n\times n)$ matrix of
	orthonormal eigenvectors, $\mathbf{V^{\prime}V=I_{N}}$, \ we obtain an
	estimate of the $r$ common factors as
	\begin{equation}
		\widehat{\mathbf{F}}_{t}=\mathbf{D}_{r}^{-1}\mathbf{V}_{r}^{\prime}%
		\mathbf{x}_{t}\text{,}%
	\end{equation}
	with $\mathbf{D}$ being the matrix containing the square root of the ordered
	eigenvalues.
	For completeness, we also discuss the relationship of FMMDE with the factor model introduced in \cite{lam2011estimation}, which can be regarded as an analog for FMMDE dealing with the case of serially linearly dependent factors. Such a model is based on the eigendecomposition of the matrix 
	\begin{equation}
		\mathbf{\mathcal{L}}_{k_{0}}=\sum_{j=1}^{k_{0}}\mathbf{cov}\left( \mathbf{x}_{t},\mathbf{x}_{t-j}\right) 
		\mathbf{cov}\left( \mathbf{x}_{t},\mathbf{x}_{t-j}\right)^{\prime},
	\end{equation}
	where $\mathbf{cov}\left( \mathbf{x}_{t},\mathbf{x}_{t-j}\right) $ is a $\left(n \times n\right)$ matrix with $\left(i,h\right)$-th entry being equal to $\text{cov}\left( x_{i,t},x_{h,t-j}\right)$.
	In particular, $\mathbf{\mathcal{L}}_{k_{0}}$, again a real, symmetric, and positive semidefinite matrix, can be considered as a linear analog of the matrix $\Gamma_{k_{0}}$ used in FMMDE. Less stringent assumptions for the error term are being made in this case, as this is defined to be a white noise sequence. In particular, the model offers a way to retrieve a decomposition between a set of serially correlated factors and a white noise idiosyncratic component. Estimation of the factor loading space works analogously, being based upon the eigenvectors of $\mathcal{L}_{k_{0}}$. The main difference between the models lies in the serial dependence structure characterizing the factors. In particular, while LYB requires factors to exhibit some degree of autocorrelation, this is not the case for FMMDE, as these can show some form of (nonlinear) dependence in the mean even when they are a white noise sequence. This is reflected in the nature of the matrices employed for estimating the models: the cumulative linear matrix $\mathbf{\mathcal{L}}_{k_{0}}$ encodes the linear dependence relationships, whereas cumulative MDDM more generally characterizes conditional mean independence.

	\subsection{Simulated Forecasts}
	\noindent We provide an example showing the forecasting advantage granted by FMMDE when factors are characterized by nonlinear serial dependence. For this reason, we consider again the nonlinear DGP seen in Section \ref{sec:factorSelection}. For completeness, we additionally consider the linear DGP defined in Example 5 in the original paper.
  In particular, the process $w_{t}$ generating the factors is assumed to follow a linear MA(1).
	\begin{equation}
		w_{t}    =0.2z_{t-1}+z_{t},z_{t}\sim N\left(  0,1\right).
	\end{equation}
	where
	$\boldsymbol{\zeta}\mathbf{_{t}}$ is a random sample of $N\left(
	0,\mathbf{\Sigma}\right)  $ and is independent of $\mathbf{F}_{t}$, where
	$\mathbf{\Sigma=}\left(  \sigma_{i,j}\right)  _{i,j=1}^{n}$ is
	defined exactly as in Equation \ref{sigma}.
	
	\noindent Assuming that the exact number of factors is known a priori (so that no factor selection procedure is involved), we evaluate each model's one-step-ahead forecast performance. Predictions are obtained for each $j$-th element in the vector $\mathbf{x_{t}}$ using a simple equation of the form 
	\begin{equation}
		\widehat{x}^{m}_{j,T+1}=\sum_{i}\widehat{\beta }%
		_{i}\widehat{F}^{m}_{i,T},
	\end{equation}
	where the correct number of underlying factors, indexed at $i$ and assumed to be known a priori, is employed. The superscript $m$ stands for the model for which the factor component $\widehat{\mathbf{F}}_{t}$ is estimated. We use the first $(T-1)$ observations for each simulated sample to estimate the underlying factor structure; this is then used to obtain the one-step ahead forecasts. For each model, $m$, forecast errors obtained for the individual series are then averaged as follows 
	\begin{equation}
		MSFE_{S}^{m}=n^{-1}\sum_{j=1}^{n}\left( \widehat{x}_{j,T+1}^{m}-x_{j,T+1}\right)^{2}
	\end{equation}
	where the subscript $S$ indicates that we are considering $S$-th simulated sample. Forecast results, reported in Table \ref{tab:Tab1SimulLinear} are expressed in terms of the ratio 
	\begin{equation}
		rMSFE=\sum_{s=1}^{1000}MSFE_{S}^{FMMDE}/\sum_{s=1}^{1000}MSFE_{S}^{m},
	\end{equation}
	where $m \in \left\{SW,LYB\right\}$, and results are averaged over 1000 simulated samples. In particular, it emerges how both the SW and LYB models have a predictive advantage over FMMDE when the factor DGP is assumed to be linear. Varying the number of variables $n$ or sample size $T$ does not have a relevant influence on the results. The table highlights the benefits of utilizing FMMDE in the presence of significant nonlinearities in the underlying factor structure. In such cases, the FMMDE outperforms the SW and LYB models. This advantage becomes more pronounced with larger values of $T$ and $n$.

	\section{Empirical Analysis \label{section:EmpAna}}
	\noindent This section aims to assess the predictive capabilities of FMMDE equipped with the factor selection procedure based on the sequential testing procedure (hereafter denoted as FMMDE$_{ST}$) on the macroeconomic aggregates from the\cite{mccracken2016fred} dataset. The performance of FMMDE$_{ST}$ is compared to that of the same model when factor selection is performed according to the eigenvalue ratio methodology (hereafter FMMDE$_{\lambda}$). The two models, SW and LYB, are also included in the analysis: the optimal number of factors is determined following \cite{bai2002determining} $IC_{p2}$ criterion in the case of SW model, whereas the eigenvalue ratio methodology is used in the case of LYB model (denoted as LYB$_{\lambda}$). All results are compared to those of a univariate autoregressive model, which is used as a benchmark for all empirical comparisons.

	\subsection{Data Description and Forecasting Procedure\label{section:forecastEQ}}
	
	\noindent We consider an $n$ dimensional stationary process $\mathbf{x}_{t} = (x_{1t},\dots,x_{nt})$  forming a large dataset of economic indicators. Let $y_{t}$ be the variable to be predicted, part of the same set of variables: we are interested in forecasting $y_{t}$ at horizon $h$ by using all the information made available by $\mathbf{x}_{t}$. The dataset consists of monthly observations on 123 U.S. macroeconomic time series observed from January 1959 to December 2019.
	The time series are grouped into eight categories listed in the Appendix. As reported in \cite{mccracken2016fred}, the series are all transformed to be stationary by taking first or second differences, logarithms, or first or second differences of logarithms. After transforming the series, the available data range from February 1962 to December 2019, for a total of $T = 695$ observations. Let $y^{h}_{t+h}$ denote the variables to be predicted in a $h$-period ahead forecast. Following \cite{stock2002macroeconomic}, when predicting real activity variables, $y^{h}_{t+h}$ is defined as the $h$-period growth at an annual rate. As an example, let $IP_t$ the level of the Industrial Production index we obtain
	\begin{equation*}
		y^{h}_{t+h} =\frac{1200}{h} \text{ln}\left(\frac{IP_{t+h}}{IP_{t}}\right),
	\end{equation*}
where $y_t = 1200*\text{ln}(IP_t/IP_{t-1})$. When predicting the nominal price and wage series, $y_{t+h}$ is defined as the $h$-period growth of monthly changes. As an example, let $CPI_t$ be the level of the Consumer Price Index; we obtain
	\begin{equation*}
		y^{h}_{t+h} = \frac{1200}{h} \text{ln}\left(\frac{CPI_{t+h}}{CPI_{t}}\right) - 1200 \text{ln} \left(\frac{CPI_{t}}{CPI_{t-1}}\right).
	\end{equation*}
	where $y_t = 1200*\Delta \text{ln}(CPI_t/CPI_{t-1})$. The forecasts are obtained through a pseudo-real-time forecasting procedure, adopting a rolling window scheme for estimation and prediction. Specifically, we opt for a rolling 10-year window.\footnote{
		At the time we started the exercise, the dataset used referred to the vintage dated March 2021. We decided to limit our analysis to December 2019 because we intended to measure the predictive capabilities of FMMDE without interfering with COVID-SARS19 pandemic effects.} Based on the estimated factors, we implement the following Diffusion Index forecasting equation 
	\begin{eqnarray}
		\widehat{y}_{t+h}^{h,m} &=& \hat{f} \left(\mathbf{z}_{t}; \hat{\boldsymbol{\vartheta}}\right) \notag \\
		&=& \widehat{\alpha }_{h}+\sum_{i=1}^{r}\widehat{\beta }%
		_{hi}\widehat{F}_{it}^{m}+\sum_{j=1}^{p}\widehat{\gamma }_{hj}y_{t-j+1},
		\label{eq:fore}
	\end{eqnarray}
	where $m$ stands for the for model for which the factor component $\widehat{\mathbf{F}}_{t}$ is estimated, so that $m \in \{\text{SW}, \text{FMMDE}_{ST}, \text{FMMDE}_{\lambda},\text{LYB}_{\lambda}\}$, $\mathbf{z}_{t} = \left(\widehat{F}_{1t}^{m},\dots,\widehat{F}_{rt}^{m}, y_{t}, \dots y_{t-p+1}\right)'$
	and $\boldsymbol{\vartheta}$ is a $q$-dimensional vector of real parameters, $\hat{\boldsymbol{\vartheta}} = (\widehat{\alpha }_{h}, \widehat{\beta }%
	_{h1}, \dots, \widehat{\beta }%
	_{hr}, \widehat{\gamma }_{h1}, \dots, \widehat{\gamma }_{hp} )$. We select the order $p$ of the autoregressive component by Bayesian Information Criterion (BIC), with $0<p\leqslant 12$. As mentioned, forecasting results obtained using diffusion indexes are compared to those obtained from a univariate autoregression, the lag order $p$ being analogously estimated by BIC, with $0<p\leqslant 12$. 
	
	\noindent The number of factors to be employed in the prediction equation is determined at each iteration of the forecasting exercise, contextually to the estimation of the latent factor space. However, since $k_{0}$  can influence the rank of matrix $\mathbf{\Gamma}_{k_0}$ and $\mathbf{\mathcal{L}}_{k_{0}}$, for FMMDE and LYB respectively, the Sequential Testing and Eigenvalue Ratio selection procedures are repeated for different values of $k_{0}$ selecting the optimal value through a cross-validation strategy tailored to the case of time series.

\noindent 	The main idea behind cross-validation is to use a sample of past predicted values, referred to as validation set, to perform the selection of the value $k_{0}$ to be employed in the proper out of sample prediction. In particular, given a discrete set of possible values for $k_{0}$, we select the parameter value that minimizes the mean squared forecast error computed over the set of predictions included in the validation set. This value, denoted as $k^{*}_{0}$, will be used to obtain the first prediction outside the validation set. In particular, let $k_0 \in \{1,\dots,20\}$ be the discrete set over which to choose $k^{*}_0$. We indicate validation and estimation sets as $\mathcal{I}_{\upsilon,k_0}^{\mathcal{V}}$ and  $\mathcal{I}_{\omega}^{\mathcal{E}}$, respectively. Moreover, $\omega = 119$ will be the size of the estimation window throughout the analysis, while $\nu=49$ is the size of the validation window. 
	The procedure is based on a rolling estimation and validation window and starts by estimating the model with the first 119 observations, namely $\mathcal{I}_{\omega}^{\mathcal{E}} = \left\{\mathbf{z}_{\tau},y_{\tau}\right\}_{\tau=Feb:62}^{Dec:71}$, which produces the forecast $\hat{y}^{h,m}_{Dec:71+h,k_0}$. Subsequently, the observation $y_{Jan:72}$ is included in the estimation sample, namely $\mathcal{I}_{\omega}^{\mathcal{E}} = \left\{\mathbf{z}_{\tau},y_{\tau}\right\}_{\tau=Mar:62}^{Jan:72}$, and the model is estimated again to get the forecast $\hat{y}_{Jan:72+h,k_0}$. The process is repeated until we have a prediction for all $49-h$ out-of-sample observations that will form the first validation set, $\mathcal{I}_{\upsilon,k_0}^{\mathcal{V}} = \left\{y_{\tau},\hat{y}^{h}_{\tau}\right\}_{\tau=Jan:72+h}^{Dec:75}$. The cross-validation mean squared forecast error (CVMSE) is then computed to select the best value for $k_0$, namely 
	$$k^{*}_{0} = \underset{k_0}{\operatorname{argmin}} \; CVMSE(k_0),$$ where 
	$$CVMSE({k_0}) = (49-h)^{-1} \sum_{\tau = Jan:72+h}^{Dec:75} \left(y_{\tau}- \hat{y}^{h,m}_{\tau,k_0}\right)^{2},$$ then the selected $k_{0}^{*}$ is used to forecast the observation $y_{Dec:75+h}$, namely $\hat{y}_{Dec:75+h,k_{0}^{*}}$. 
	To construct the forecast for observation $y_{Jan:76+h}$, we update the validation set, namely $\mathcal{I}_{\upsilon}^{\mathcal{V}} = \left\{y_{\tau},\hat{y}^{h}_{\tau,k_0}\right\}_{\tau=Feb:72+h}^{Jan:76}$, obtaining a new value for $k^{*}_0$ which is used to calculate the second forecast, namely $\hat{y}_{Jan:76+h,k_{0}^{*}}$. The process is repeated until the end of the sample is reached, producing a total of 516 forecasts, ranging from January 1976 to January 2019 for $h=1$, June 1976 to June 2019 for $h=6$, and December 1976 to December 2019 for $h=12$. The predictive performance of the models is finally  evaluated in terms of mean square
	forecast error, $MSFE_{h}^{m}$, defined as
	\[
	MSFE_{h,k^{\star}_{0}}^{m}=\left(T_{1}-T_{0}\right)^{-1}\sum_{\tau=T_{0}+h}^{T_{1}}\left(
	y_{\tau}-\hat{y}_{\tau,k^{\star}_{0}}^{h,m}\right)  ^{2},
	\] where $T_{0}$ is the first point in time for out-of-sample evaluation and
	$T_{1}$ is the last point in time for which we compute MSFE at horizon $h$. 

	\section{Empirical Results\label{section:empirical results}}
\noindent 	The results of the empirical exercise are reported in Tables \ref{tab:singleSeries} and \ref{Tab3:SelSerRec}. In table \ref{tab:singleSeries}, the results pertain to some key variables of the US economy, while table \ref{Tab3:SelSerRec} presents aggregated results considering all the series comprising the 8 sectors of the dataset. The  models are compared at horizons $h=1$, $6$, and $12$, in terms of the ratio%
	\begin{equation*}
		rMSFE_{h}^{m}=MSFE_{h}^{m}/MSFE_{h}^{AR},
	\end{equation*}
	where rMSFE is the relative mean square forecast error with respect  to AR(BIC) model and $ m \in \{FMMDE_{ST}, FMMDE_{\lambda}, LYB, SW\}$. 

\subsection{Results for Target Variables}
\noindent  This section reports the  results of the forecasting analysis for the Industrial Production index (IP), Unemployment Rate (UR), Nonfarm Employment (PAYEMS), Consumer Price index inflation (CPI), and Core CPI inflation (CoCPI). 

\vspace{0.3cm}

\noindent  By analyzing the results obtained for single variables, reported in Table \ref{tab:singleSeries}-Panel A, we observe an improvement of the FMMDE$_{ST}$ procedure compared to FMMDE$_{\lambda}$ for the variables IP, UR, and PAYEMS. This result is confirmed across all forecast horizons. However, if we look at the results for the nominal variables, i.e., CPI and cCPI, in this case, FMMDE$_{\lambda}$ has an advantage, especially for $h=12$.  When comparing different factor models (martingale difference and SW) they both show a clear improvement over the forecasts obtained with an AR model, especially for the variables IP, UR, and PAYEMS. However, factor models tend to be less accurate in forecasting the US nominal variables. These results are in line with previous findings by \cite{eickmeier2008successful}. The previous results are further supported by the \cite{diebold1995comparing} test, as shown in Table \ref{tab:singleSeries} - Panel B. The test confirms that both FMMDE$_{ST}$ and SW significantly outperform AR(BIC) in terms of predictive ability, though this is less evident for FMMDE$_{\lambda}$. In the case of the CPI, we cannot reject the null hypothesis of equal predictive ability for all diffusion index models considered, except for FMMDE$_{\lambda}$ when $h=12$.

 \vspace{0.3cm}
	
\begin{figure}[ht]
		
		\centering
		\begin{subfigure}[b]{0.49\textwidth}
			\centering
			\caption{$IP,h=1$}
			\includegraphics[width=\textwidth]{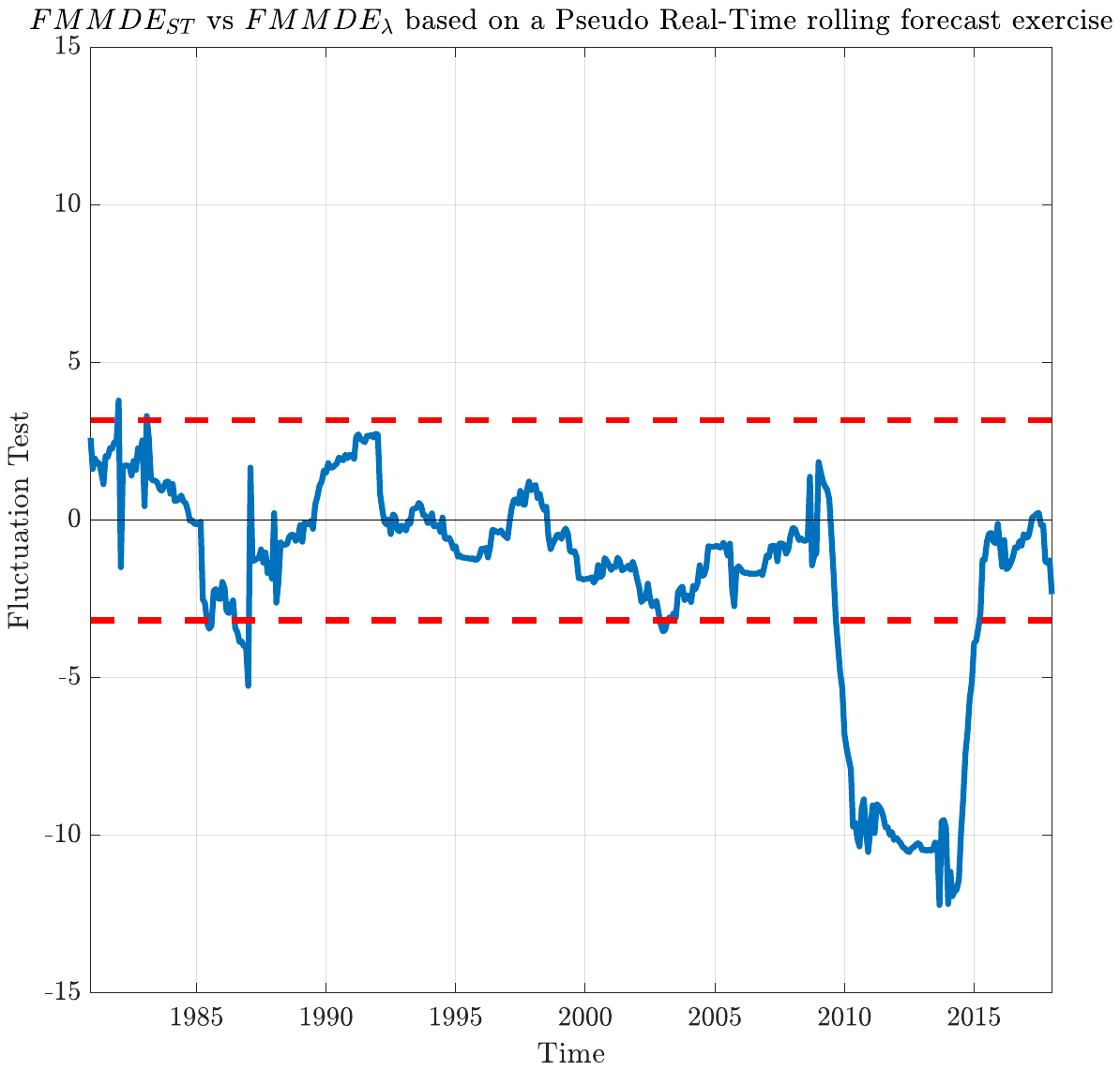}
			
		\end{subfigure}
		\begin{subfigure}[b]{0.49\textwidth}
			\centering
			\caption{$IP,h=12$}
			\includegraphics[width=\textwidth]{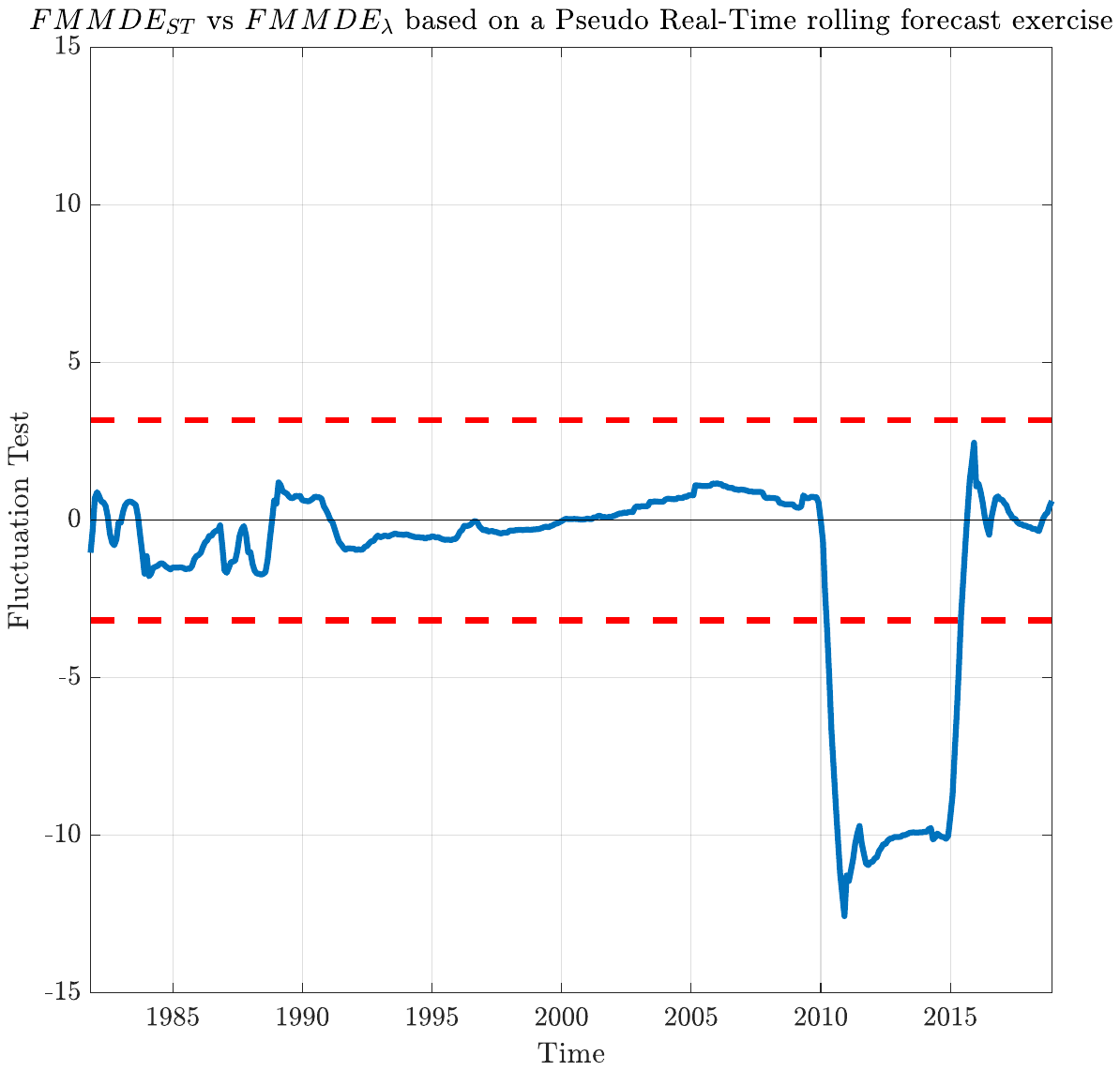}
		\end{subfigure}
		\par
		\vspace{22pt}
		\centering
		\begin{subfigure}[b]{0.49\textwidth}
			\centering
			\caption{$CPI,h=1$}
			\includegraphics[width=\textwidth]{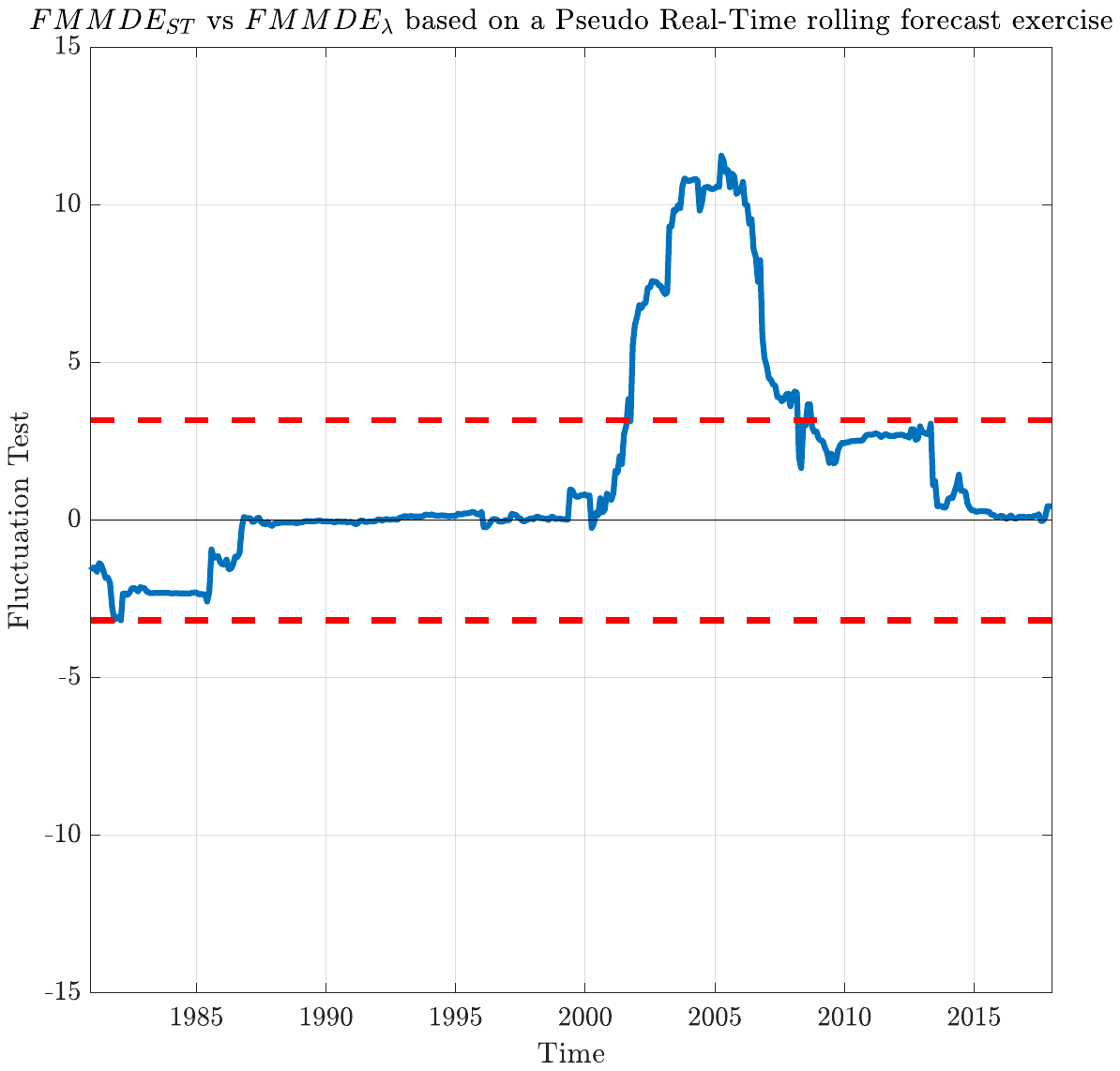}
			
		\end{subfigure}
		\begin{subfigure}[b]{0.49\textwidth}
			\centering
			\caption{$CPI,h=12$}
			\includegraphics[width=\textwidth]{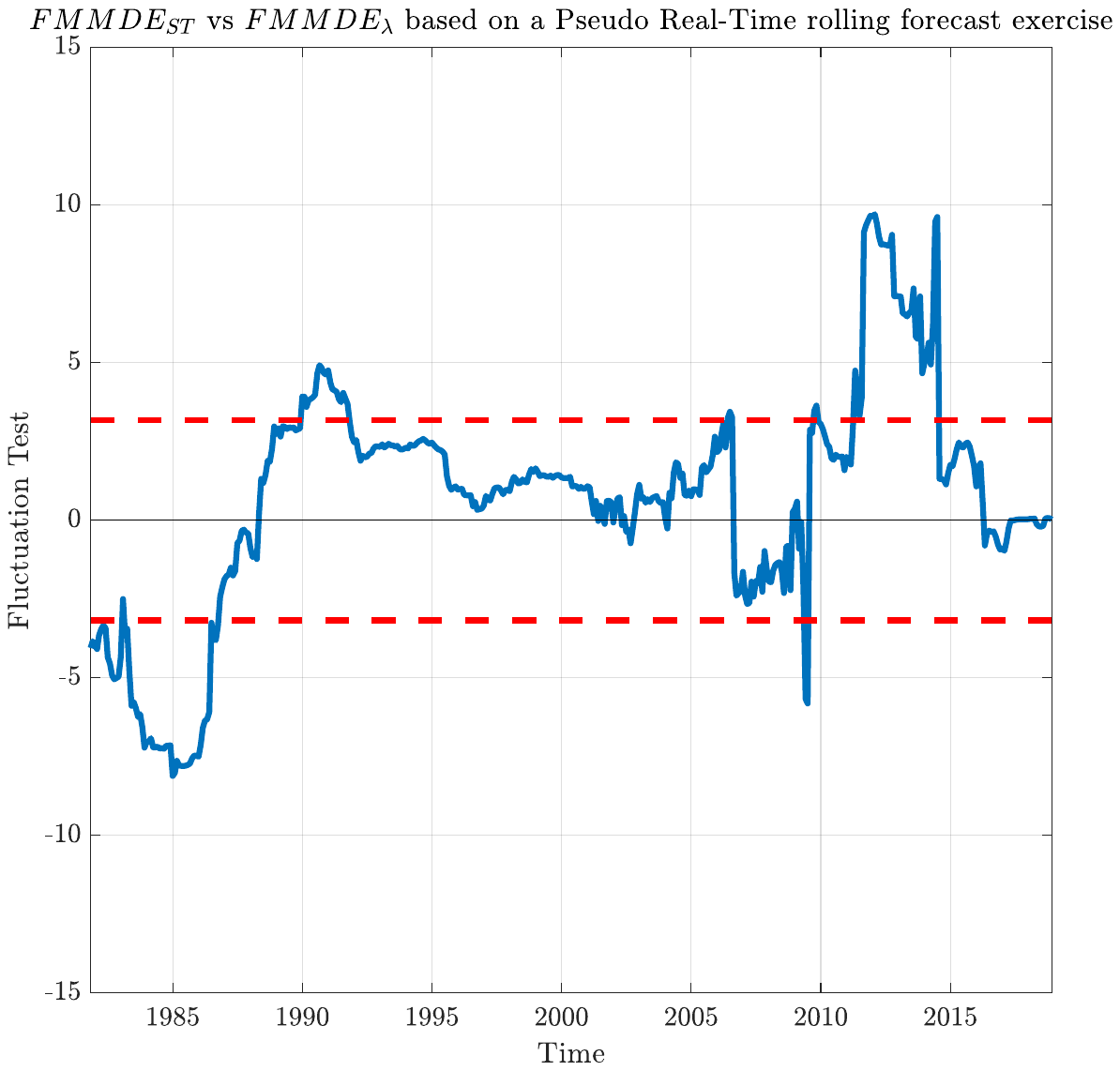}
		\end{subfigure}
		\par
		\vspace{22pt}
		\centering
		\begin{subfigure}[b]{0.49\textwidth}
			\centering
			\caption{$UNRATE,h=1$}
			\includegraphics[width=\textwidth]{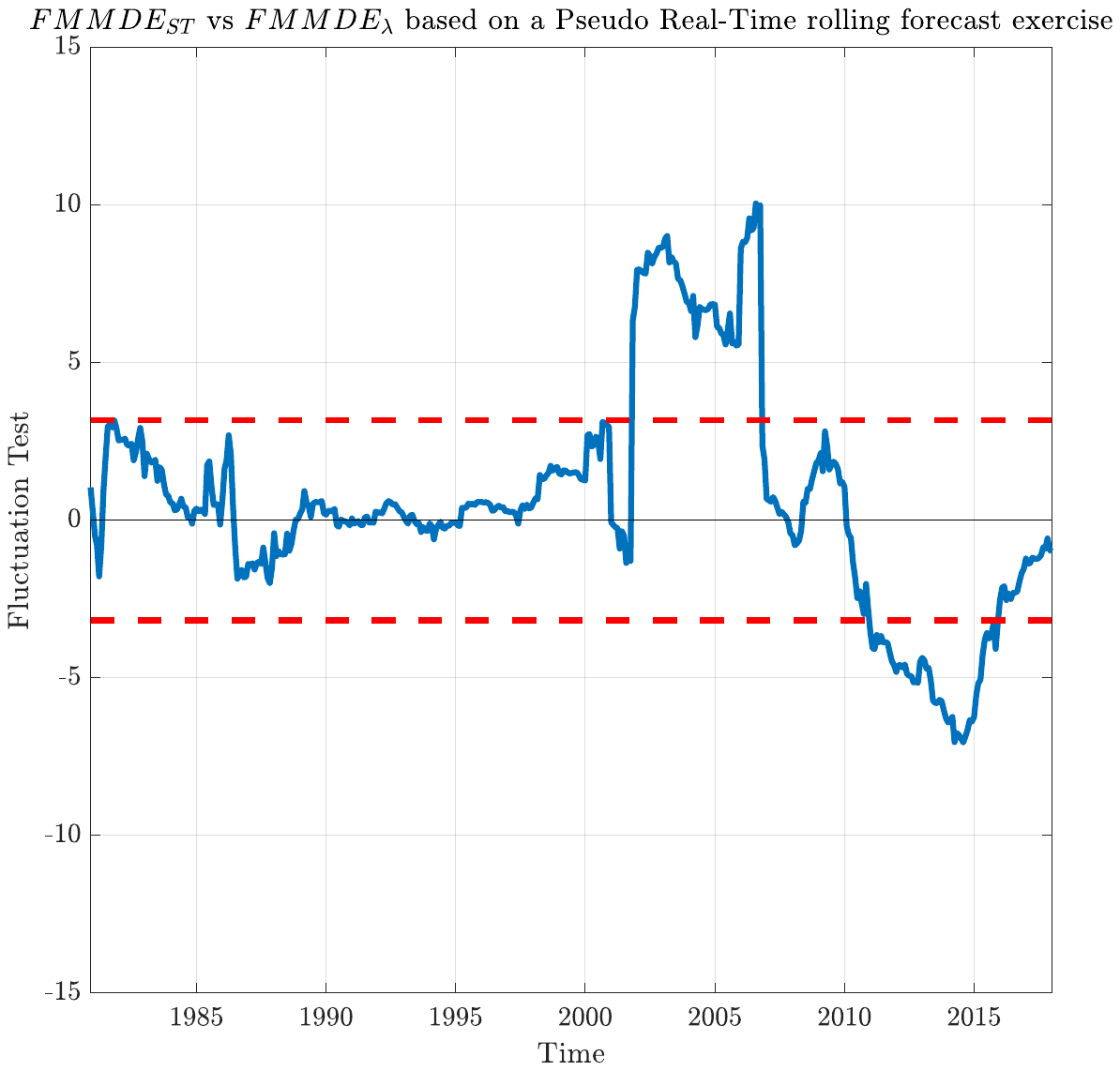}
		\end{subfigure}
		\begin{subfigure}[b]{0.49\textwidth}
			\centering
			\caption{$UNRATE,h=12$}
			\includegraphics[width=\textwidth]{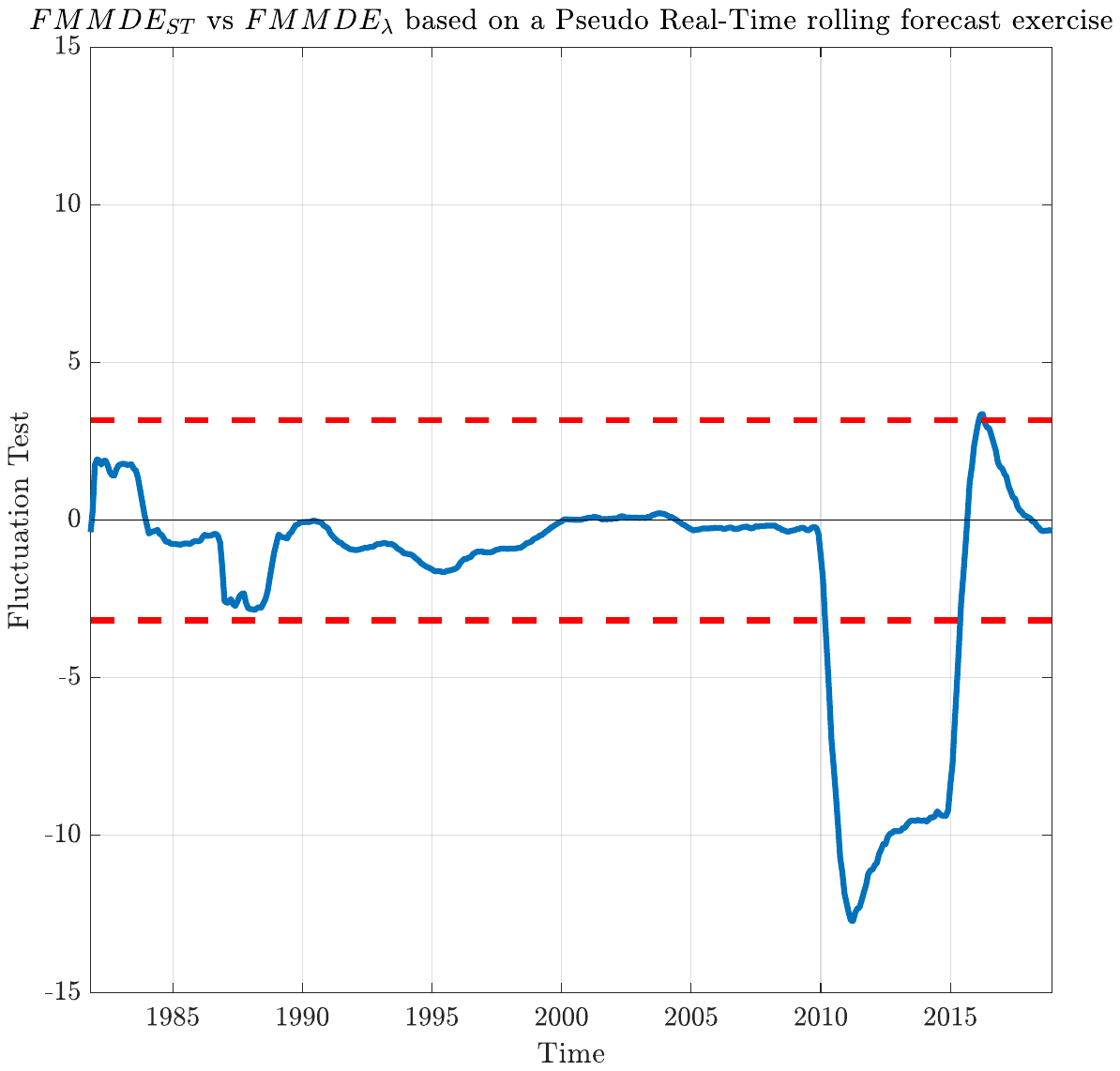}
		\end{subfigure}
		\caption{ Fluctuation test (IP), (CPI), (UNRATE):  $FMMDE_{ST}$ versus $FMMDE_{\lambda}$ model, $h=1,12$. Notes: Fluctuation test statistic: solid. 5$\%$ critical value: dotted. If the solid is below the dotted (red) line the
			$FMMDE_{ST}$ method is significantly (better) than the $FMMDE_{\lambda}$ method, and vice versa.}%
		\label{fig:gr_st_vs_er}%
	\end{figure}

\begin{figure}[ht]
		
		\centering
		\begin{subfigure}[b]{0.49\textwidth}
			\centering
			\caption{$IP,h=1$}
			\includegraphics[width=\textwidth]{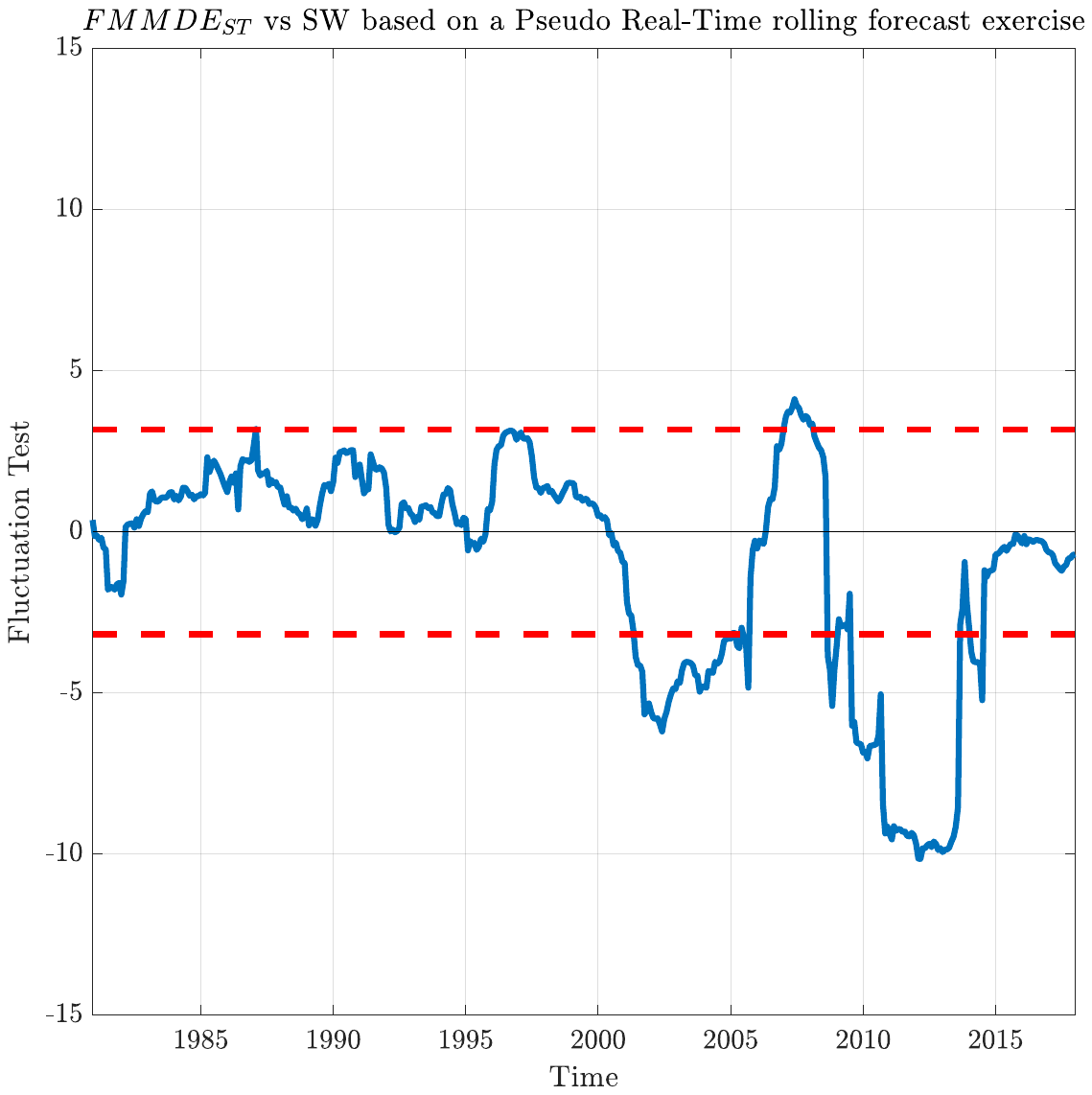}
			
		\end{subfigure}
		\begin{subfigure}[b]{0.49\textwidth}
			\centering
			\caption{$IP,h=12$}
			\includegraphics[width=\textwidth]{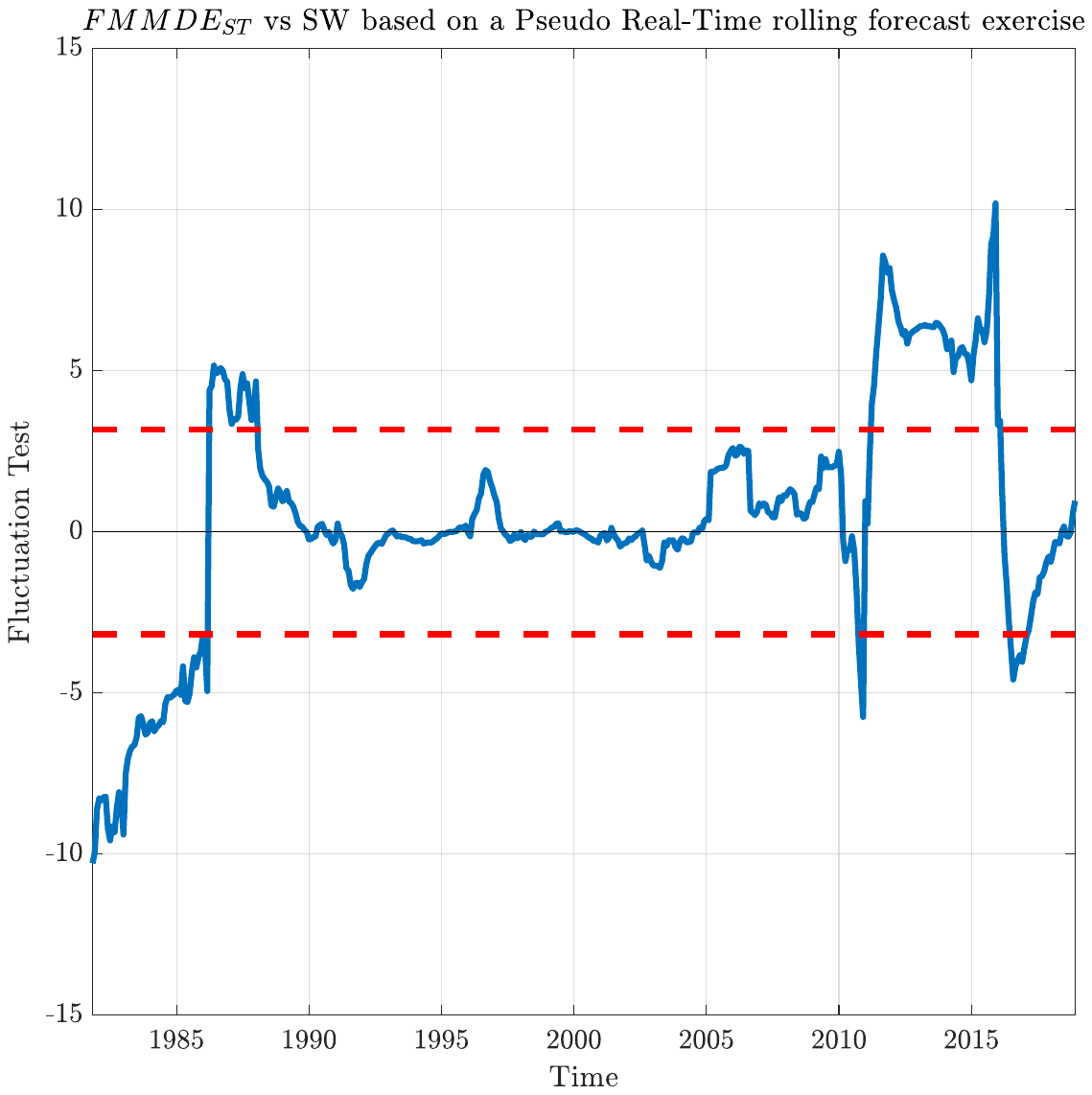}
		\end{subfigure}
		\par
		\vspace{22pt}
		\centering
		\begin{subfigure}[b]{0.49\textwidth}
			\centering
			\caption{$CPI,h=1$}
			\includegraphics[width=\textwidth]{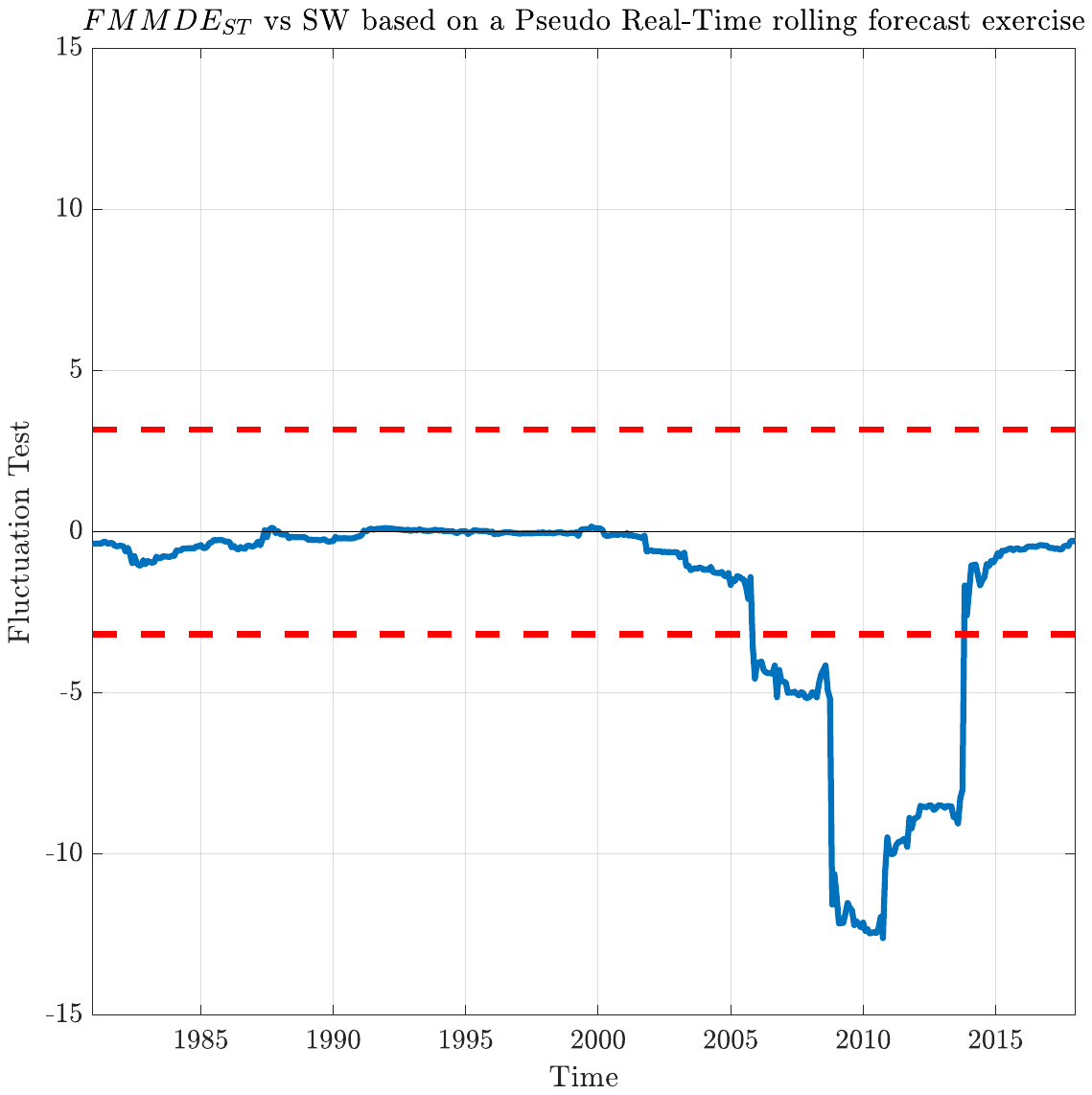}
			
		\end{subfigure}
		\begin{subfigure}[b]{0.49\textwidth}
			\centering
			\caption{$CPI,h=12$}
			\includegraphics[width=\textwidth]{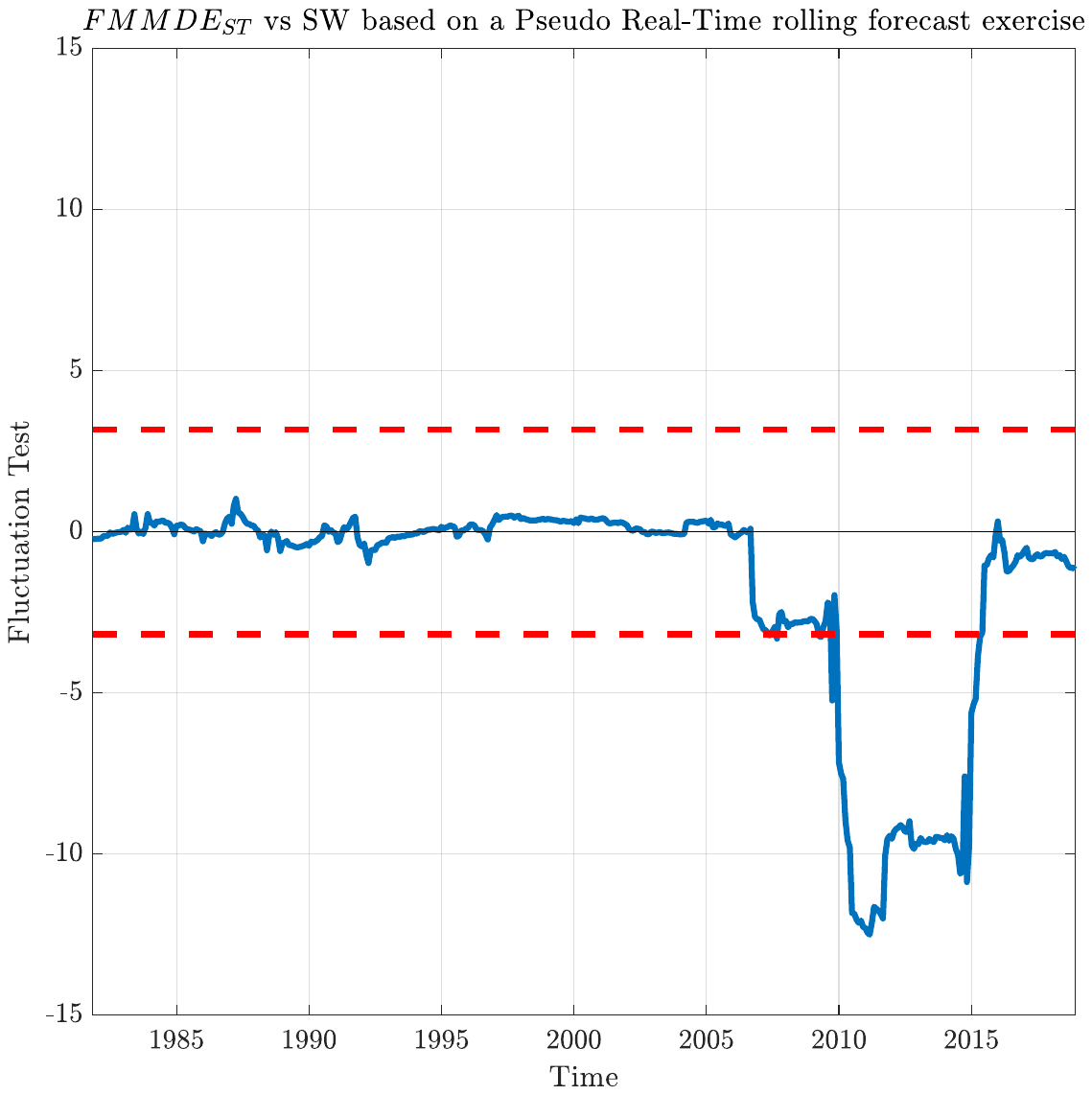}
		\end{subfigure}
		\par
		\vspace{22pt}
		\centering
		\begin{subfigure}[b]{0.49\textwidth}
			\centering
			\caption{$UNRATE,h=1$}
			\includegraphics[width=\textwidth]{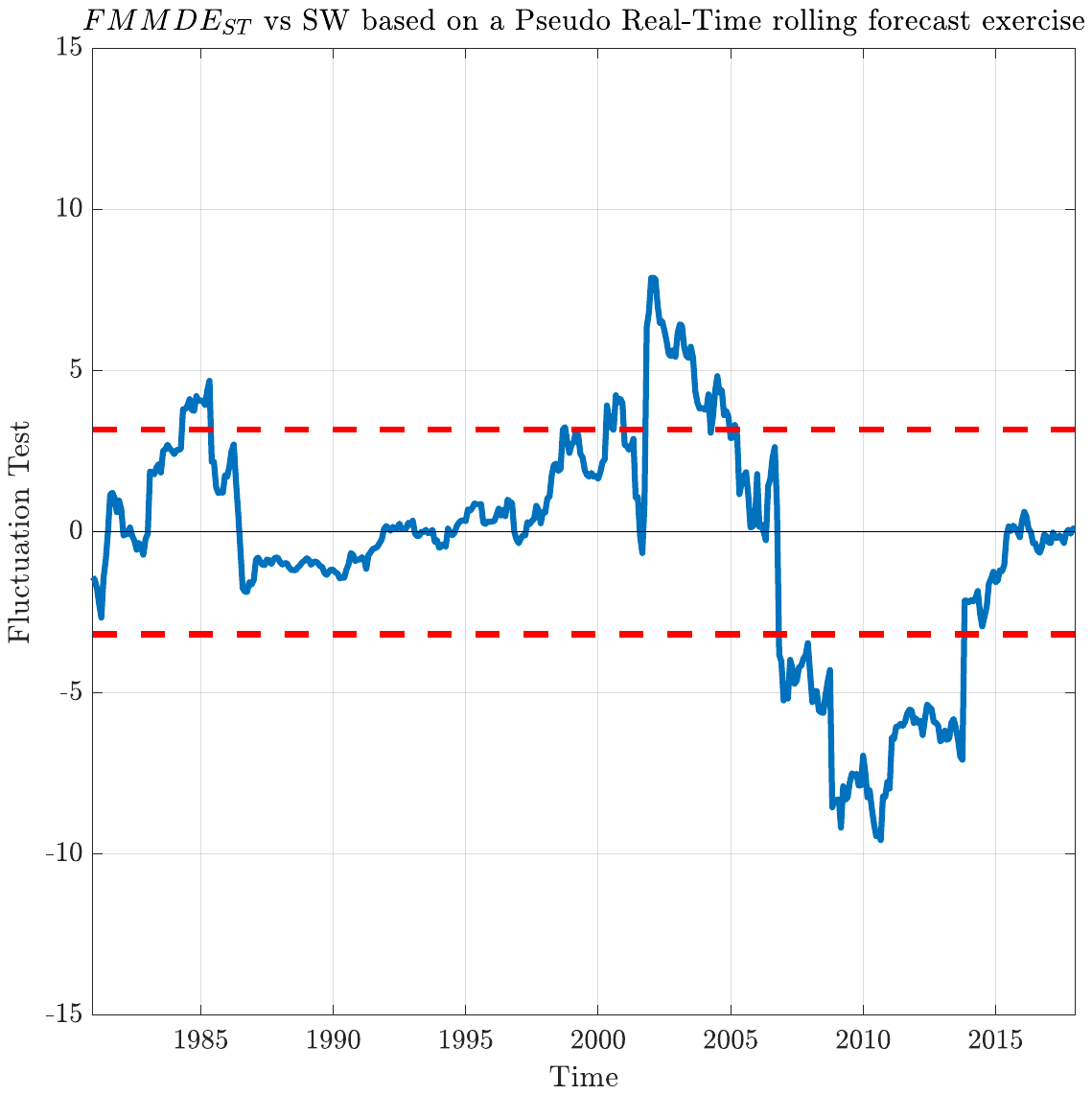}
		\end{subfigure}
		\begin{subfigure}[b]{0.49\textwidth}
			\centering
			\caption{$UNRATE,h=12$}
			\includegraphics[width=\textwidth]{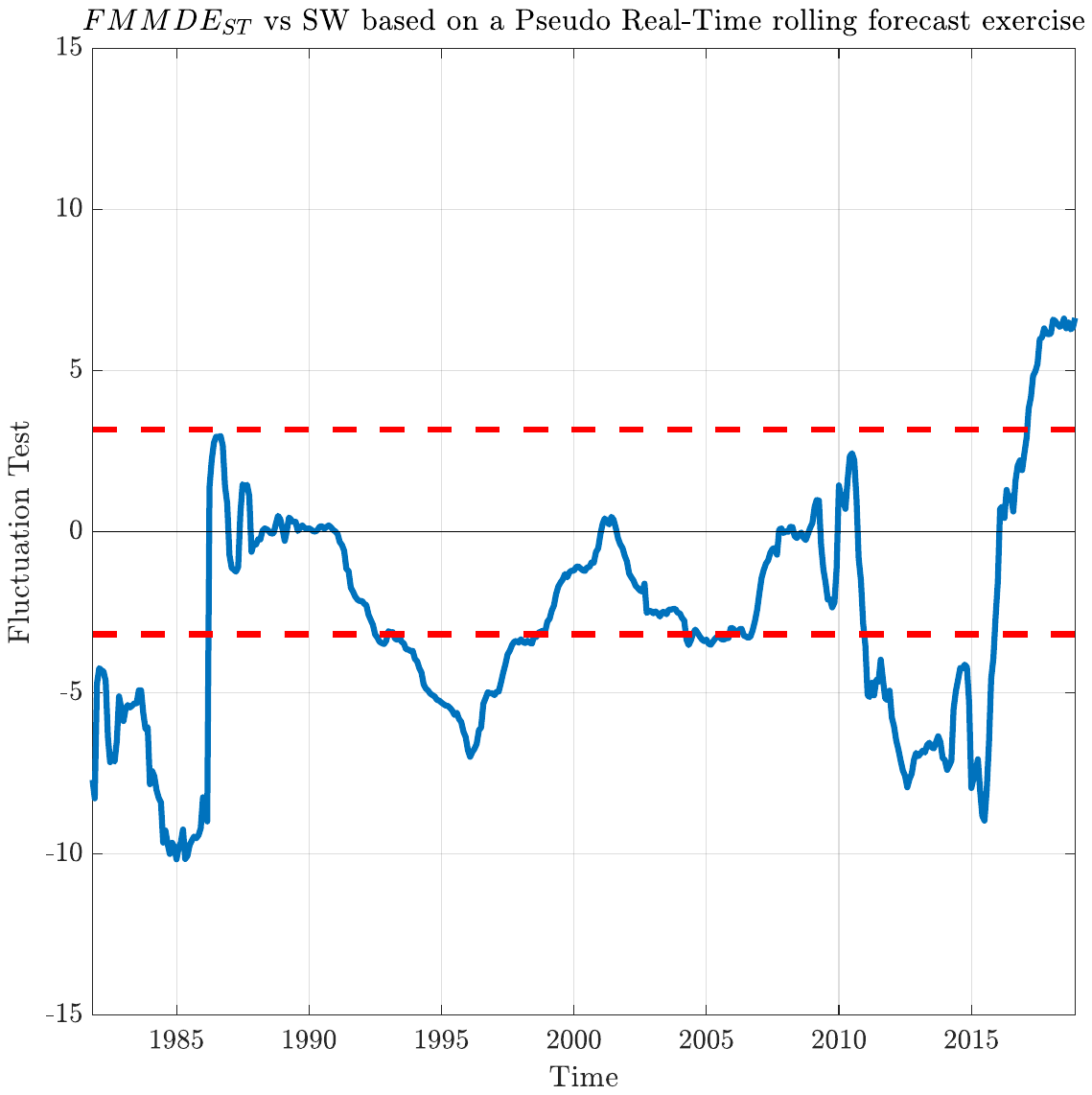}
		\end{subfigure}
		\caption{ Fluctuation test (IP), (CPI), (UNRATE):  $FMMDE_{ST}$ versus $SW$ model, $h=1,12$. Notes: Fluctuation test statistic: solid. 5$\%$ critical value: dotted. If the solid is below the dotted (red) line the
			$FMMDE_{ST}$ method is significantly (better) than the second, and vice versa.}%
		\label{fig:gr_st_vs_sw}%
	\end{figure}

	\noindent  As further evidence, to address potential instabilities that might affect the Diebold-Mariano test, we adopt the fluctuation test procedure introduced by \cite{giacomini2010forecast} (hereafter GR) to locally evaluate the forecasting performance among the procedures at specific points in time.\footnote{This approach is detailed in \cite{giacomini2010forecast} as a simple way to test against the null of equal local performance of two forecasting methods.} The evolution of the relationship between the forecast performance of FMMDE$_{ST}$ and that of FMMDE$_{\lambda}$, and SW is analyzed in detail in Figure \ref{fig:gr_st_vs_er} and \ref{fig:gr_st_vs_sw} respectively.\footnote{We have not included the comparison with LYB as it exhibits very similar performance with FMMDE$_{\lambda}$} The two figures display results obtained for the fluctuation test for the time series of IP, UNRATE, and CPI at horizons $h=1,12$. The solid line represents the graph of the difference between the square forecast error obtained for the two models, normalized by its estimated standard deviation and smoothed by a centered moving average of length $L = 61$, with the coefficients equal to $1/L$. The zero horizontal line is indicative of the two models having equal performance, while the dotted lines indicate the 5$\%$ critical values, so that FMMDE$_{ST}$ outperforms (underperforms) the other models locally, at the 5$\%$ significance level when the solid line is below (above) the lower (upper) dashed line.\footnote{Because the moving averages are of length 61 and centered, the first and last 30 values are not computed or graphed.}

\noindent  As illustrated in Figure \ref{fig:gr_st_vs_er}, for variables such as IP and UNRATE, FMMDE$_{ST}$ and FMMDE$_{\lambda}$ exhibit comparable performance throughout most of the evaluation period. However, during the years following the Great Recession and subsequent recovery (starting from the second half of 2009 until around the end of 2015), FMMDE$_{ST}$ has the best performance. Regarding CPI, FMMDE$_{\lambda}$ has the best results in the mid-2000s for $h = 1$ and during the years between 2011 and 2015 for $h = 12$. Regarding the comparison with the SW model, as illustrated in Figure \ref{fig:gr_st_vs_sw}, FMMDE$_{ST}$ exhibits a significant advantage over the SW model during the period that coincides with the Great Recession for both forecast horizons, with the exception of IPI when $h=12$.

		\subsection{Results for groups of variables}
	
	\noindent We report the distribution of MSFE relative to AR for a specific group of variables in Table \ref{Tab3:SelSerRec}. In particular, we consider seven distinct groups: Output \& Income (19 series in total), Labor Market (31 series), Housing Market (10 series), Consumption, Business Inventories \& Orders, Money \& Credit (12 series), Interest \& Exchange Rates (21 series) and Prices (16 series). The results previously observed for individual series are confirmed. FMMDE$_{ST}$ has an advantage over FMMDE$_{\lambda}$ in the prediction of real economic variables (roughly represented by the first four groups in the dataset); this is partly balanced by the higher accuracy displayed by FMMDE$_{\lambda}$ in forecasting nominal variables (represented by the last three groups).

 \noindent  When instead the different procedures are compared, we notice that FMMDE$_{ST}$ delivers more accurate forecasts than FMMDE$_\lambda$ for more than half of the series for the first four groups at all considered horizons. This result extends to more than three-quarters of the series for $h=6,12$. No significant differences emerge in the performance of FMMDE$_\lambda$ and LYB$_\lambda$. The performance of FMMDE$_{ST}$ appears to align with that of the SW model in the prediction of the first four groups, even though FMMDE$_{ST}$ has an advantage for $h=1$, more prominently in the case of the fourth group. All the diffusion index models provide a better forecasting performance than the AR model for more than half of the series in the first four groups, except for FMMDE$_\lambda$ and LYB$_\lambda$ when it comes to longer horizons ($h=12$). On the other hand, FMMDE$_{ST}$ is less performing compared to FMMDE$_\lambda$ and LYB for the groups: Money \& Credit, Interest \& Exchange Rates, and Prices.

	\subsection{Key Take away}
\begin{figure}[ht]
		
		\centering
		\centering
		\begin{subfigure}[b]{1\textwidth}
			\centering
			\includegraphics[width=5in]{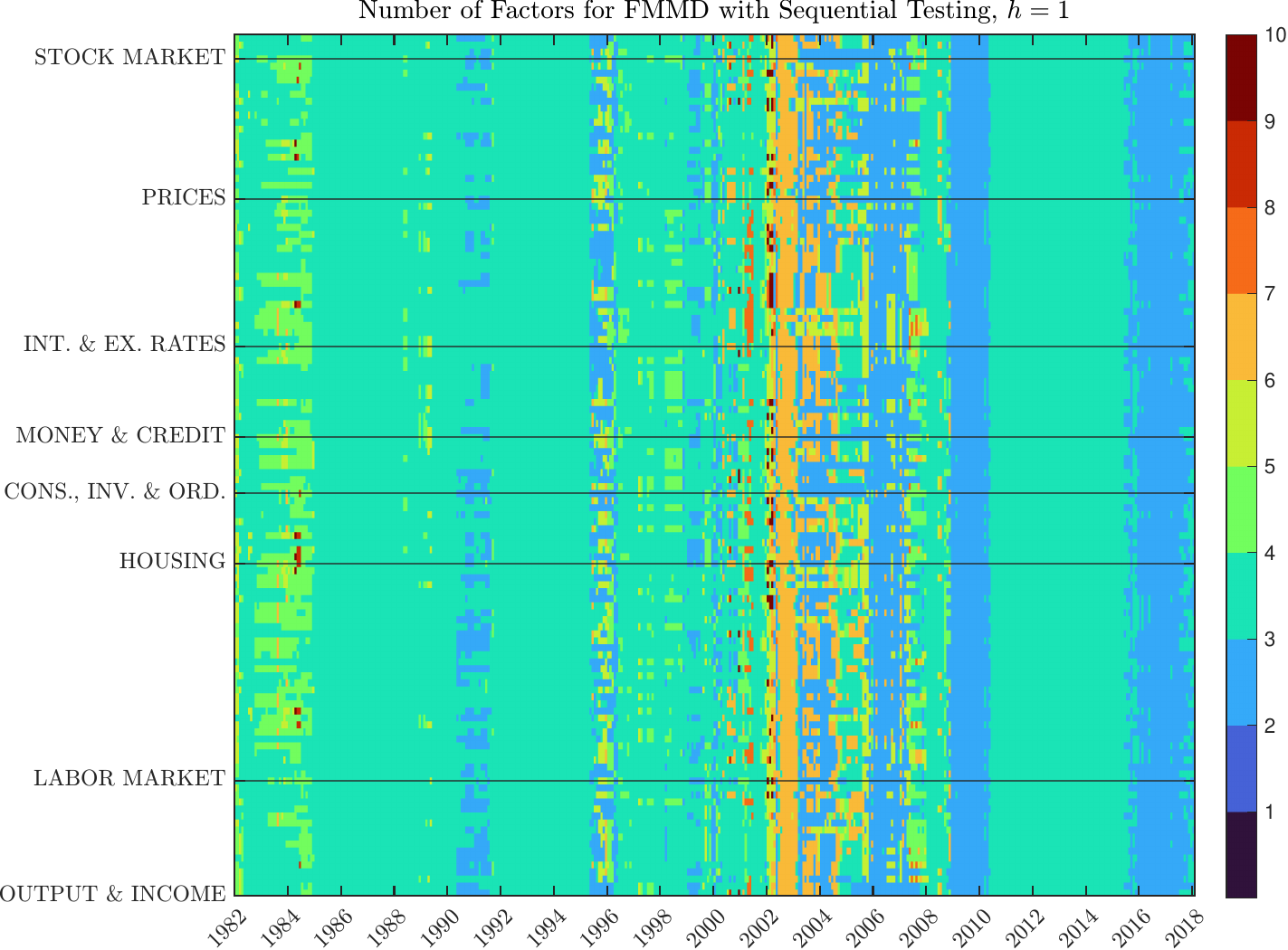}
			
		\end{subfigure}
		\par
		\vspace{22pt}
		\centering
		\begin{subfigure}[b]{1\textwidth}
			\centering
			\includegraphics[width=5in]{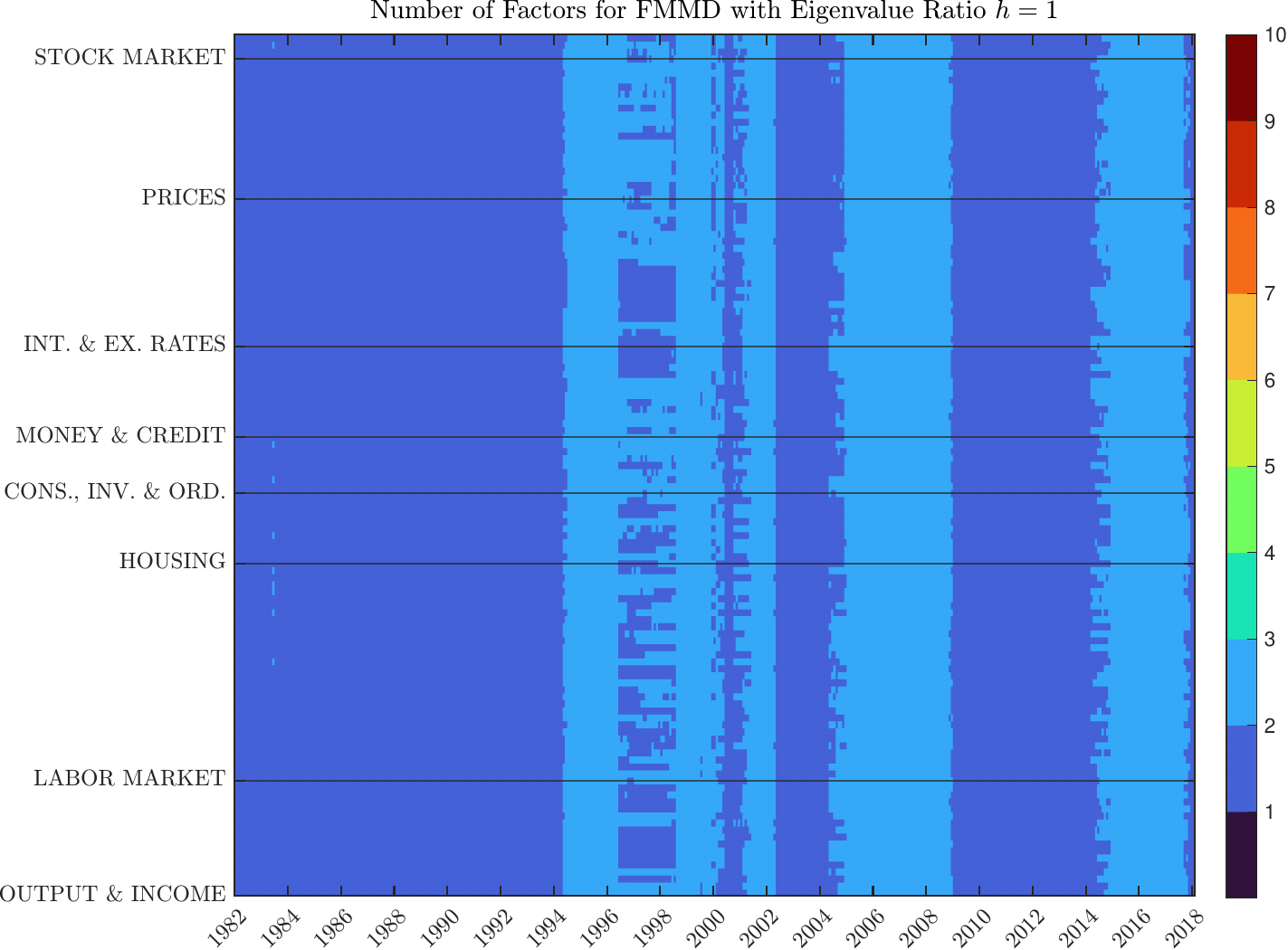}
		\end{subfigure}
		\caption{The figure shows the selection of factors over time for the Sequential Testing procedure (top figure) and Eigenvalue Ratio (bottom figure).}%
		\label{fig:fac_sel}%
	\end{figure}

	\noindent First, we observe that our selection procedure is much less conservative than the eigenvalue ratio methodology in selecting the number of factors, which explains the results obtained with our forecasting experiment see Figure \ref{fig:fac_sel}. Selecting a larger number of factors gives a clear advantage in predicting real variables. This result, however, comes at the cost of a lower accuracy in the prediction of nominal variables.

\noindent Second, the cross-validation selection of the lags used by the matrix $\widehat{\boldmath{\Gamma}}_{k0}$ influences the factor selection procedure. We can deduce this from the cross-sectional variability observable on the vertical axis of the plot for any fixed time. However, such an effect is limited to specific time intervals, and the degree of variability is limited, as predicting individual variables leads us, in general, to consider latent factor spaces of similar dimensions.

	\section*{Conclusions}
	\noindent In this paper, we have examined the predictive performance of factor models with martingale difference error (FMMDE). As a primary contribution, we have introduced a novel procedure for selecting the optimal number of factors for FMMDE. Through simulation experiments, we show the good finite sample properties of our proposed methodology. Furthermore, we have evaluated the predictive performance of FMMDE in comparison to other models, such as SW and LYB, using real data. Specifically, we have considered a large dataset of monthly macroeconomic and financial series for the US economy (FRED dataset). The results indicate that, overall, the competing methods exhibit comparable forecasting performance. However, for real activity variables, such as Industrial Production and Unemployment, our selection procedure offers advantages to the FMMDE method, particularly for medium to long-term forecasting horizons.
	
	\clearpage
	
	\begin{table}[ptbh!]
		
		\renewcommand*{\arraystretch}{1.1} \centering
		\vspace{0.05cm} \scalebox{0.7}{

\begin{tabular}{cc|ccccccc|cccccc|}
				
				\cmidrule{3-15} 
				&       &       &       & \multicolumn{4}{c}{\textbf{Sequential Test}}      &       &       & \multicolumn{4}{c}{\textbf{Eigenvalue Ratio}}     &  \\
				\cmidrule{5-8}\cmidrule{11-14}      \diagbox{$r$}{$\theta$}    &       &       &       & \multicolumn{1}{l}{$r$} & \multicolumn{1}{l}{$0.5r$} & \multicolumn{1}{l}{$3r$} & \multicolumn{1}{l}{       $5r$} &       &       & \multicolumn{1}{l}{$r$} & \multicolumn{1}{l}{$0.5r$} & \multicolumn{1}{l}{$3r$} & \multicolumn{1}{l}{       $5r$} &  \\
				\hline \hline
				&       &       &       &       &       &       &       &       &       &       &       &       &       &  \\
				&       & \multicolumn{1}{l}{$DGP_{1}$} &       & \multicolumn{1}{c}{\cellcolor[rgb]{ .751,  .751,  .751}0.86} & \multicolumn{1}{c}{\cellcolor[rgb]{ .751,  .751,  .751}0.87} & \multicolumn{1}{c}{\cellcolor[rgb]{ .751,  .751,  .751}0.80} & \multicolumn{1}{c}{\cellcolor[rgb]{ .751,  .751,  .751}0.73} &       &       & \multicolumn{1}{c}{0.82} & \multicolumn{1}{c}{0.77} & \multicolumn{1}{c}{0.53} & \multicolumn{1}{c}{0.40} &  \\
				\multicolumn{2}{c|}{2} & \multicolumn{1}{l}{$DGP_{2}$} &       & \multicolumn{1}{c}{\cellcolor[rgb]{ .751,  .751,  .751}0.87} & \multicolumn{1}{c}{\cellcolor[rgb]{ .751,  .751,  .751}0.86} & \multicolumn{1}{c}{\cellcolor[rgb]{ .751,  .751,  .751}0.73} & \multicolumn{1}{c}{\cellcolor[rgb]{ .751,  .751,  .751}0.64} &       &       & \multicolumn{1}{c}{0.80} & \multicolumn{1}{c}{0.71} & \multicolumn{1}{c}{0.47} & \multicolumn{1}{c}{0.33} &  \\
				&       & \multicolumn{1}{l}{$DGP_{3}$} &       & \multicolumn{1}{c}{\cellcolor[rgb]{ .751,  .751,  .751}0.84} & \multicolumn{1}{c}{\cellcolor[rgb]{ .751,  .751,  .751}0.69} & \multicolumn{1}{c}{\cellcolor[rgb]{ .751,  .751,  .751}0.24} & \multicolumn{1}{c}{\cellcolor[rgb]{ .751,  .751,  .751}0.10} &       &       & \multicolumn{1}{c}{0.57} & \multicolumn{1}{c}{0.37} & \multicolumn{1}{c}{0.16} & \multicolumn{1}{c}{0.13} &  \\
\end{tabular}}
		\caption{Empirical probability of estimating the true number of factors: $n=50$, $T=50$ (upper part of the table), and $n=50$,$T=100$ (lower part of the table).}%
		\label{tab:simu1n50}%
		
	\end{table}
	
	\clearpage

	\begin{table}[ptbh!]
		\renewcommand*{\arraystretch}{1.1} \centering
		\vspace{0.05cm} \scalebox{0.7}{ 
			
			\begin{tabular}{cc|ccccccc|cccccc|}
				\cmidrule{3-15}
				&       &       &       & \multicolumn{4}{c}{\textbf{Sequential Test}}      &       &       & \multicolumn{4}{c}{\textbf{Eigenvalue Ratio}}     &  \\
				\cmidrule{5-8}\cmidrule{11-14}      \diagbox{$r$}{$\theta$}    &       &       &       & \multicolumn{1}{l}{$r$} & \multicolumn{1}{l}{$0.5r$} & \multicolumn{1}{l}{$3r$} & \multicolumn{1}{l}{       $5r$} &       &       & \multicolumn{1}{l}{$r$} & \multicolumn{1}{l}{$0.5r$} & \multicolumn{1}{l}{$3r$} & \multicolumn{1}{l}{       $5r$} &  \\
				\hline \hline
				&       &       &       &       &       &       &       &       &       &       &       &       &       &  \\
				&       & \multicolumn{1}{l}{$DGP_{1}$} &       & 0.87 & \multicolumn{1}{c}{\cellcolor[rgb]{ .751,  .751,  .751}0.87} & \multicolumn{1}{c}{\cellcolor[rgb]{ .751,  .751,  .751}0.85} & \multicolumn{1}{c}{\cellcolor[rgb]{ .751,  .751,  .751}0.80} &       &       & \multicolumn{1}{c}{\cellcolor[rgb]{ .751,  .751,  .751}0.90} & \multicolumn{1}{c}{0.81} & \multicolumn{1}{c}{0.66} & \multicolumn{1}{c}{0.57} &  \\
				\multicolumn{2}{c|}{2} & \multicolumn{1}{l}{$DGP_{2}$} &       & \multicolumn{1}{c}{\cellcolor[rgb]{ .751,  .751,  .751}0.89} & \multicolumn{1}{c}{\cellcolor[rgb]{ .751,  .751,  .751}0.87} & \multicolumn{1}{c}{\cellcolor[rgb]{ .751,  .751,  .751}0.81} & \multicolumn{1}{c}{\cellcolor[rgb]{ .751,  .751,  .751}0.75} &       &       & \multicolumn{1}{c}{0.83} & \multicolumn{1}{c}{0.74} & \multicolumn{1}{c}{0.54} & \multicolumn{1}{c}{0.41} &  \\
				&       & \multicolumn{1}{l}{$DGP_{3}$} &       & \multicolumn{1}{c}{\cellcolor[rgb]{ .751,  .751,  .751}0.83} & \multicolumn{1}{c}{\cellcolor[rgb]{ .751,  .751,  .751}0.81} & \multicolumn{1}{c}{\cellcolor[rgb]{ .751,  .751,  .751}0.45} & \multicolumn{1}{c}{\cellcolor[rgb]{ .751,  .751,  .751}0.18} &       &       & \multicolumn{1}{c}{0.67} & \multicolumn{1}{c}{0.50} & \multicolumn{1}{c}{0.21} & \multicolumn{1}{c}{0.12} &  \\
\end{tabular}}
		\caption{Empirical probability of estimating the true number of factors: $n=100$, $T=50$ (upper part of the table), and $n=100$,$T=100$ (lower part of the table).}%
		\label{tab:simu2n100}%
	\end{table}
	\clearpage
	
	\begin{table}[ptbh!]
		\renewcommand*{\arraystretch}{1.1} \centering
		\vspace{0.05cm} \scalebox{0.7}{ 
			\begin{tabular}{cc|ccccccc|cccccc|}
				\cmidrule{3-15}
				&       &       &       & \multicolumn{4}{c}{\textbf{Sequential Test}}      &       &       & \multicolumn{4}{c}{\textbf{Eigenvalue Ratio}}     &  \\
				\cmidrule{5-8}\cmidrule{11-14}      \diagbox{$r$}{$\theta$}    &       &       &       & \multicolumn{1}{l}{$r$} & \multicolumn{1}{l}{$0.5r$} & \multicolumn{1}{l}{$3r$} & \multicolumn{1}{l}{       $5r$} &       &       & \multicolumn{1}{l}{$r$} & \multicolumn{1}{l}{$0.5r$} & \multicolumn{1}{l}{$3r$} & \multicolumn{1}{l}{       $5r$} &  \\
				\hline \hline
				&       &       &       &       &       &       &       &       &       &       &       &       &       &  \\
				&       & \multicolumn{1}{l}{$DGP_{1}$} &       & 0.96  & 0.97  & \cellcolor[rgb]{ .751,  .751,  .751}0.95 & \cellcolor[rgb]{ .751,  .751,  .751}0.94 &       &       & \cellcolor[rgb]{ .751,  .751,  .751}0.99 & \cellcolor[rgb]{ .751,  .751,  .751}0.98 & 0.95  & 0.88  &  \\
				\multicolumn{2}{c|}{2} & \multicolumn{1}{l}{$DGP_{2}$} &       & 0.97  & \cellcolor[rgb]{ .751,  .751,  .751}0.97 & \cellcolor[rgb]{ .751,  .751,  .751}0.96 & \cellcolor[rgb]{ .751,  .751,  .751}0.94 &       &       & \cellcolor[rgb]{ .751,  .751,  .751}0.99 & 0.96  & 0.88  & 0.81  &  \\
				&       & \multicolumn{1}{l}{$DGP_{3}$} &       & 0.94  & \cellcolor[rgb]{ .751,  .751,  .751}0.95 & \cellcolor[rgb]{ .751,  .751,  .751}0.92 & \cellcolor[rgb]{ .751,  .751,  .751}0.74 &       &       & \cellcolor[rgb]{ .751,  .751,  .751}0.96 & 0.86  & 0.55  & 0.29  &  \\
				&       &       &       &       &       &       &       &       &       &       &       &       &       &  \\
				&       & \multicolumn{1}{l}{$DGP_{1}$} &       & 0.96  & 0.96  & \cellcolor[rgb]{ .751,  .751,  .751}0.95 & \cellcolor[rgb]{ .751,  .751,  .751}0.94 &       &       & \cellcolor[rgb]{ .751,  .751,  .751}0.99 & \cellcolor[rgb]{ .751,  .751,  .751}0.98 & 0.87  & 0.74  &  \\
\end{tabular}}
		\caption{Empirical probability of estimating the true number of factors: $n=200$,$T=100$ (upper part of the table) and $n=100$,$T=200$ (lower part of the table).}
		\label{tab:simu3n100}
	\end{table}

	\clearpage
	
	\bigskip
	\bigskip
	\begin{table}[htbp]
		\renewcommand*{\arraystretch}{1.1} \centering
		\vspace{0.05cm} \scalebox{0.7}{ 
			\begin{tabular}{cc|cc|ccccc|cccccc|}
				\toprule\toprule
				&       &          &    & \multicolumn{4}{r}{\textbf{Sequential Test}}      &        & &     \multicolumn{4}{c}{\textbf{Eigenvalue Ratio}}   &   \\
				\cmidrule{3-4}\cmidrule{6-8}\cmidrule{11-14}     $k_{0}$     &       &     \diagbox{$T$}{$n$}  &    &   & \multicolumn{1}{c}{$50$} & \multicolumn{1}{c}{$100$} & \multicolumn{1}{c}{$200$}  &       &  &     & \multicolumn{1}{l}{$50$} & \multicolumn{1}{c}{$100$} & \multicolumn{1}{c}{$200$}  &  \\
				\hline \hline
				&       &       &       &       &       &       &       &    &   &      & &       &       &               \\
				&       & \multicolumn{1}{c}{$50$} &   &    & 0.07  & 0.05  &0.03  &     &  &       & \cellcolor[rgb]{ .751,  .751,  .751}0.71 & \cellcolor[rgb]{ .751,  .751,  .751}0.75 & \cellcolor[rgb]{ .751,  .751,  .751}0.75  &   \\
				\multicolumn{2}{c|}{1} & \multicolumn{1}{c}{$100$} & &      & 0.27  & 0.22 & 0.19 &       &&       & \cellcolor[rgb]{ .751,  .751,  .751}0.82 &\cellcolor[rgb]{ .751,  .751,  .751}0.89  & \cellcolor[rgb]{ .751,  .751,  .751}0.94  &  \\
				&       & \multicolumn{1}{c}{$200$} &  &     &  0.68 & 0.69 &  0.76 &     &  &       & \cellcolor[rgb]{ .751,  .751,  .751}0.95 & \cellcolor[rgb]{ .751,  .751,  .751}0.98 & \cellcolor[rgb]{ .751,  .751,  .751}0.99 &  \\

				&       &       &       &       &       &    &   &       &       &       &  &     &       &              \\
				&       & \multicolumn{1}{c}{$50$} &   &    &\cellcolor[rgb]{ .751,  .751,  .751} 0.64  & \cellcolor[rgb]{ .751,  .751,  .751}0.82  &\cellcolor[rgb]{ .751,  .751,  .751}0.90  &     &  &       & 0.43 & 0.48 & 0.58  &   \\
\end{tabular}}
		\caption{Empirical probability of estimating the true number of factors: Example 6 .}
		\label{tab:simEx6}
	\end{table}

	\bigskip
	\bigskip
	\bigskip
	\bigskip
	\bigskip
	\bigskip
	\bigskip
	\bigskip
	\bigskip
	\bigskip
	\bigskip

	\begin{table}[htbp]
		\centering \centering
		\vspace{0.05cm}
		\scalebox{0.7}{	\begin{tabular}{r|rrr|rrr|lll|lll|}
				\toprule\toprule
				\multicolumn{1}{r}{} & \multicolumn{6}{c}{SW Model}                  & \multicolumn{6}{c}{LYB Model} \\
				\cmidrule{2-13}        & \multicolumn{3}{c}{Linear Factors} & \multicolumn{3}{c|}{Nonlinear Factors} & \multicolumn{3}{c}{Linear Factors} & \multicolumn{3}{c|}{Nonlinear Factors} \\
				\cmidrule{1-13}
				\diagbox{$n$}{$T$} & 200   & 500   & \multicolumn{1}{c|}{1000} & 200   & 500   & \multicolumn{1}{c|}{1000} & \multicolumn{1}{c}{200} & \multicolumn{1}{c}{500} & \multicolumn{1}{c|}{1000} & \multicolumn{1}{c}{200} & \multicolumn{1}{c}{500} & \multicolumn{1}{c|}{1000} \\
				\cmidrule{1-13}
				100   & 1.039 & 1.032 & 1.041 & 1.010 & \cellcolor[rgb]{ .751,  .751,  .751}0.974 & \cellcolor[rgb]{ .751,  .751,  .751}0.943 &1.010 &1.014 &1.020 & 1.006 & \cellcolor[rgb]{ .751,  .751,  .751}0.952 & \cellcolor[rgb]{ .751,  .751,  .751}0.915 \\
				
				300   & 1.051 & 1.022 & 1.026 & \cellcolor[rgb]{ .751,  .751,  .751}0.978 & \cellcolor[rgb]{ .751,  .751,  .751}0.955 & \cellcolor[rgb]{ .751,  .751,  .751}0.961 & 1.019 & 1.010 & 1.014 & \cellcolor[rgb]{ .751,  .751,  .751}0.966 & \cellcolor[rgb]{ .751,  .751,  .751}0.947 & \cellcolor[rgb]{ .751,  .751,  .751}0.933 \\
				
				500   & 1.040 & 1.057 & 1.051 & \cellcolor[rgb]{ .751,  .751,  .751}0.975 & \cellcolor[rgb]{ .751,  .751,  .751}0.926 & \cellcolor[rgb]{ .751,  .751,  .751}0.927 & 1.019 & 1.027 & 1.034 & \cellcolor[rgb]{ .751,  .751,  .751}0.938 & \cellcolor[rgb]{ .751,  .751,  .751}0.885 & \cellcolor[rgb]{ .751,  .751,  .751}0.882 \\
				\bottomrule
			\end{tabular}%
		}
		\caption{Ratio of the average one step ahead mean square forecast errors (MSFE) computed for the two models over 1000 simulations, FMMDE (Numerator), SW and LYB model (Denominator)}%
		\label{tab:Tab1SimulLinear}%
	\end{table}%

	\begin{table}[htbp]
		\centering
		\caption{MSFE relative to AR, for relevant macroeconomic aggregates}
		\scalebox{0.70}{\begin{tabular}{rrrrrrrrrrrrrrrr}
				\toprule
				\toprule
				&       &       &       &       &       &       &       &       &       &       &       &       &       &       &  \\
				& \multicolumn{15}{c}{\textbf{PANEL A: Relative MSE }} \\
				\cmidrule{2-16}          &       &       &       &       &       &       &       &       &       &       &       &       &       &       &  \\
				& \multicolumn{3}{c}{\boldmath{}\textbf{FMMDE$_{ST}$ vs AR}\unboldmath{}} &       & \multicolumn{3}{c}{\textbf{SW vs AR}} &       & \multicolumn{3}{c}{\boldmath{}\textbf{FMMDE$_{\lambda}$ vs AR}\unboldmath{}} &       & \multicolumn{3}{c}{\boldmath{}\textbf{LYB$_{\lambda}$ vs AR}\unboldmath{}} \\
				\cmidrule{2-4}\cmidrule{6-8}\cmidrule{10-12}\cmidrule{14-16}          & \multicolumn{1}{c}{$h=1$} & \multicolumn{1}{c}{$h=6$} & \multicolumn{1}{c}{$h=12$} &       & \multicolumn{1}{c}{$h=1$} & \multicolumn{1}{c}{$h=6$} & \multicolumn{1}{c}{$h=12$} &       & \multicolumn{1}{c}{$h=1$} & \multicolumn{1}{c}{$h=6$} & \multicolumn{1}{c}{$h=12$} &       & \multicolumn{1}{c}{$h=1$} & \multicolumn{1}{c}{$h=6$} & \multicolumn{1}{c}{$h=12$} \\
				&       &       &       &       &       &       &       &       &       &       &       &       &       &       &  \\
				\multicolumn{1}{l}{IP} & \multicolumn{1}{c}{\cellcolor[rgb]{ .682,  .667,  .667}0.870} & \multicolumn{1}{c}{\cellcolor[rgb]{ .682,  .667,  .667}0.812} & \multicolumn{1}{c}{\cellcolor[rgb]{ .682,  .667,  .667}0.932} &       & \multicolumn{1}{c}{\cellcolor[rgb]{ .682,  .667,  .667}0.894} & \multicolumn{1}{c}{\cellcolor[rgb]{ .682,  .667,  .667}0.915} & \multicolumn{1}{c}{\cellcolor[rgb]{ .682,  .667,  .667}0.926} &       & \multicolumn{1}{c}{\cellcolor[rgb]{ .682,  .667,  .667}0.907} & \multicolumn{1}{c}{\cellcolor[rgb]{ .682,  .667,  .667}0.992} & \multicolumn{1}{c}{1.085} &       & \multicolumn{1}{c}{\cellcolor[rgb]{ .682,  .667,  .667}0.911} & \multicolumn{1}{c}{1.008} & \multicolumn{1}{c}{1.093} \\
\end{tabular}}%
		\begin{center}
			\begin{minipage}[h]{15cm}
				\begin{spacing}{1.5}
					\scriptsize \textbf{NOTE}: Panel A reports the relative MSFEs with respect to AR(BIC) at the 1-,6- and 12-month ahead forecast horizon. Entries below one, highlighted by grey areas, indicate better performance than AR(BIC). MSFEs are calculated using rolling pseudo out-of-sample forecasts over the period 1976-2019. Panel B reports the Diebold-Mariano (DM) statistics for the paired comparison of the AR(BIC) with models reported in column. Entries in bold  denote rejection of the null hypothesis of equal predictive ability.
				\end{spacing}
			\end{minipage}
		\end{center}
		\label{tab:singleSeries}
		
	\end{table}%

	\clearpage
	\begin{table}[htbp]
		\centering
		\caption{Distribution of MSFE relative to AR for a specific group of variables}
		\scalebox{0.625}{
\begin{tabular}{rrrrrrrrrrrrrrrrrr}
				\toprule
				\toprule
				&       &       &       &       &       &       &       &       &       &       &       &       &       &       &       &       &  \\
				& \multicolumn{17}{c}{\boldmath{}\textbf{PANEL A: $DEC-1975+h  : APR-2019$ FMMDE$_{ST}$}\unboldmath{}} \\
				\cmidrule{2-18}          & \multicolumn{5}{c}{$h=1$}             &       & \multicolumn{5}{c}{$h=6$}             &       & \multicolumn{5}{c}{$h=12$} \\
				\cmidrule{2-6}\cmidrule{8-12}\cmidrule{14-18}          & \multicolumn{1}{c}{10th} & \multicolumn{1}{c}{25th} & \multicolumn{1}{c}{50th} & \multicolumn{1}{c}{75th} & \multicolumn{1}{c}{90th} &       & \multicolumn{1}{c}{10th} & \multicolumn{1}{c}{25th} & \multicolumn{1}{c}{50th} & \multicolumn{1}{c}{75th} & \multicolumn{1}{c}{90th} &       & \multicolumn{1}{c}{10th} & \multicolumn{1}{c}{25th} & \multicolumn{1}{c}{50th} & \multicolumn{1}{c}{75th} & \multicolumn{1}{c}{90th} \\
				&       &       &       &       &       &       &       &       &       &       &       &       &       &       &       &       &  \\
				\multicolumn{1}{l}{OUTPUT \& INCOME} & \multicolumn{1}{c}{\cellcolor[rgb]{ .682,  .667,  .667}0.873} & \multicolumn{1}{c}{\cellcolor[rgb]{ .682,  .667,  .667}0.896} & \multicolumn{1}{c}{\cellcolor[rgb]{ .682,  .667,  .667}0.955} & \multicolumn{1}{c}{\cellcolor[rgb]{ .682,  .667,  .667}0.990} & \multicolumn{1}{c}{1.026} &       & \multicolumn{1}{c}{\cellcolor[rgb]{ .682,  .667,  .667}0.786} & \multicolumn{1}{c}{\cellcolor[rgb]{ .682,  .667,  .667}0.819} & \multicolumn{1}{c}{\cellcolor[rgb]{ .682,  .667,  .667}0.857} & \multicolumn{1}{c}{1.008} & \multicolumn{1}{c}{1.076} &       & \multicolumn{1}{c}{\cellcolor[rgb]{ .682,  .667,  .667}0.831} & \multicolumn{1}{c}{\cellcolor[rgb]{ .682,  .667,  .667}0.920} & \multicolumn{1}{c}{\cellcolor[rgb]{ .682,  .667,  .667}0.935} & \multicolumn{1}{c}{1.055} & \multicolumn{1}{c}{1.072} \\
\bottomrule
		\end{tabular}}%
		\par
		\begin{center}
			\begin{minipage}[h]{15.6cm}
				\begin{spacing}{1.5}
					\scriptsize \textbf{NOTE}: Entries are percentiles of distributions MSFEs over the 123 variables being forecasted, by series, at the 1-,6- and 12-month ahead forecast horizon. All forecasts are direct. MSFEs are calculated using rolling pseudo out-of-sample forecasts over the period 1976-2019. Grey areas indicate in which percentiles a given model is performing better than AR(BIC).
				\end{spacing}
			\end{minipage}
		\end{center}
		\label{Tab3:SelSerRec}
	\end{table}
	
	\clearpage
	
	\clearpage
	
	\clearpage

 \clearpage
	\bibliography{references}

\end{document}


We report the list of series used in the forecasting exercise. We use the same mnermoni group and gtransformation as described in McCracken and Ng. In particular, the column tcode denotes the following data transformation for a series $x_t$: (1) no transformation; (2) $\Delta x_t$; (3)$\Delta^2 x_t$; (4) $ log(x_t)$; (5) $\Delta log(x_t)$; (6) $\Delta^2 log(x_t)$. (7) $\Delta (x_t/x_{t-} - 1.0)$. The FRED column gives mnemonics in FRED followed by a short description. The comparable series in Global Insight is given in the column GSI.
\footnotesize
\section*{Appendix A: List of Series}
\begin{longtable}{cllcc}
\caption{A sample long table.} \label{tab:long} \\

\hline \multicolumn{1}{c}{\textbf{ID}} & \multicolumn{1}{c}{\textbf{FRED}} & \multicolumn{1}{c}{\textbf{Description}} & \multicolumn{1}{c}{\textbf{Group}} & \multicolumn{1}{c}{\textbf{Tcode}} \\ \hline 
\endfirsthead

\multicolumn{5}{c}%
{{\bfseries \tablename\ \thetable{} -- continued from previous page}} \\
\hline \multicolumn{1}{c}{\textbf{ID}} & \multicolumn{1}{c}{\textbf{FRED}} & \multicolumn{1}{c}{\textbf{Description}} & \multicolumn{1}{c}{\textbf{Group}} & \multicolumn{1}{c}{\textbf{Tcode}} \\ \hline
\endhead

\hline \multicolumn{5}{r}{{Continued on next page}} \\ 
\endfoot

\hline \hline
\endlastfoot

        &       &       &       &  \\
    1     & RPI   & Real Personal Income & 1     & 5 \\
    2     & W875RX1 & Real personal income ex transfer receipts & 1     & 5 \\
    3     & DPCERA3M086SBEA & Real personal consumption expenditures & 4     & 5 \\
    4     & CMRMTSPLx & Real Manu.  and Trade Industries Sales & 4     & 5 \\
    5     & RETAILx & Retail and Food Services Sales & 4     & 5 \\
    6     & INDPRO & IP Index & 1     & 5 \\
    7     & IPFPNSS & IP: Final Products and Nonindustrial Supplies & 1     & 5 \\
    8     & IPFINAL & IP: Final Products (Market Group) & 1     & 5 \\
    9     & IPCONGD & IP: Consumer Goods & 1     & 5 \\
    10    & IPDCONGD & IP: Durable Consumer Goods & 1     & 5 \\
    11    & IPNCONGD & IP: Nondurable Consumer Goods & 1     & 5 \\
    12    & IPBUSEQ & IP: Business Equipment & 1     & 5 \\
    13    & IPMAT & IP: Materials & 1     & 5 \\
    14    & IPDMAT & IP: Durable Materials & 1     & 5 \\
    15    & IPNMAT & IP: Nondurable Materials & 1     & 5 \\
    16    & IPMANSICS & IP: Manufacturing (SIC) & 1     & 5 \\
    17    & IPB51222s & IP: Residential Utilities & 1     & 5 \\
    18    & IPFUELS & IP: Fuels & 1     & 5 \\
    19    & CUMFNS & Capacity Utilization:  Manufacturing & 1     & 2 \\
    20    & HWI   & Help-Wanted Index for United States & 2     & 2 \\
    21    & HWIURATIO & Ratio of Help Wanted/No.  Unemployed & 2     & 2 \\
    22    & CLF16OV & Civilian Labor Force & 2     & 5 \\
    23    & CE16OV & Civilian Employment & 2     & 5 \\
    24    & UNRATE & Civilian Unemployment Rate & 2     & 2 \\
    25    & UEMPMEAN & Average Duration of Unemployment (Weeks) & 2     & 2 \\
    26    & UEMPLT5 & Civilians Unemployed - Less Than 5 Weeks & 2     & 5 \\
    27    & UEMP5TO14 & Civilians Unemployed for 5-14 Weeks & 2     & 5 \\
    28    & UEMP15OV & Civilians Unemployed - 15 Weeks \& Over & 2     & 5 \\
    29    & UEMP15T26 & Civilians Unemployed for 15-26 Weeks & 2     & 5 \\
    30    & UEMP27OV & Civilians Unemployed for 27 Weeks and Over & 2     & 5 \\
    31    & CLAIMSx & Initial Claims & 2     & 5 \\
    32    & PAYEMS & All Employees:  Total nonfarm & 2     & 5 \\
    33    & USGOOD & All Employees:  Goods-Producing Industries & 2     & 5 \\
    34    & CES1021000001 & All Employees:  Mining and Logging:  Mining & 2     & 5 \\
    35    & USCONS & All Employees:  Construction & 2     & 5 \\
    36    & MANEMP & All Employees:  Manufacturing & 2     & 5 \\
    37    & DMANEMP & All Employees:  Durable goods & 2     & 5 \\
    38    & NDMANEMP & All Employees:  Nondurable goods & 2     & 5 \\
    39    & SRVPRD & All Employees:  Service-Providing Industries & 2     & 5 \\
    40    & USTPU & All Employees:  Trade, Transportation \& Utilities & 2     & 5 \\
    41    & USWTRADE & All Employees:  Wholesale Trade & 2     & 5 \\
    42    & USTRADE & All Employees:  Retail Trade & 2     & 5 \\
    43    & USFIRE & All Employees:  Financial Activities & 2     & 5 \\
    44    & USGOVT & All Employees:  Government & 2     & 5 \\
    45    & CES0600000007 & Avg Weekly Hours :  Goods-Producing & 2     & 1 \\
    46    & AWOTMAN & Avg Weekly Overtime Hours :  Manufacturing & 2     & 2 \\
    47    & AWHMAN & Avg Weekly Hours :  Manufacturing & 2     & 1 \\
    48    & HOUST & Housing Starts:  Total New Privately Owned & 3     & 4 \\
    49    & HOUSTNE & Housing Starts, Northeast & 3     & 4 \\
    50    & HOUSTMW & Housing Starts, Midwest & 3     & 4 \\
    51    & HOUSTS & Housing Starts, South & 3     & 4 \\
    52    & HOUSTW & Housing Starts, West & 3     & 4 \\
    53    & PERMIT & New Private Housing Permits (SAAR) & 3     & 4 \\
    54    & PERMITNE & New Private Housing Permits, Northeast (SAAR) & 3     & 4 \\
    55    & PERMITMW & New Private Housing Permits, Midwest (SAAR) & 3     & 4 \\
    56    & PERMITS & New Private Housing Permits, South (SAAR) & 3     & 4 \\
    57    & PERMITW & New Private Housing Permits, West (SAAR) & 3     & 4 \\
    58    & AMDMNOx & New Orders for Durable Goods & 4     & 5 \\
    59    & AMDMUOx & Unfilled Orders for Durable Goods & 4     & 5 \\
    60    & BUSINVx & Total Business Inventories & 4     & 5 \\
    61    & ISRATIOx & Total Business:  Inventories to Sales Ratio & 4     & 2 \\
    62    & M1SL  & M1 Money Stock & 5     & 6 \\
    63    & M2SL  & M2 Money Stock & 5     & 6 \\
    64    & M2REAL & Real M2 Money Stock & 5     & 5 \\
    65    & BOGMBASE & Monetary Base & 5     & 6 \\
    66    & TOTRESNS & Total Reserves of Depository Institutions & 5     & 6 \\
    67    & NONBORRES & Reserves Of Depository Institutions & 5     & 7 \\
    68    & BUSLOANS & Commercial and Industrial Loans & 5     & 6 \\
    69    & REALLN & Real Estate Loans at All Commercial Banks & 5     & 6 \\
    70    & NONREVSL & Total Nonrevolving Credit & 5     & 6 \\
    71    & CONSPI & Nonrevolving consumer credit to Personal Income & 5     & 2 \\
    72    & S\&P 500 & S\&Ps Common Stock Price Index: Composite & 8     & 5 \\
    73    & S\&P: indust & S\&Ps Common Stock Price Index: Industrials & 8     & 5 \\
    74    & S\&P div yield & S\&Ps Composite Common Stock: Dividend Yield & 8     & 2 \\
    75    & S\&P PE ratio & S\&Ps Composite Common Stock: Price-Earnings Ratio & 8     & 5 \\
    76    & FEDFUNDS & Effective Federal Funds Rate & 6     & 2 \\
    77    & CP3Mx & 3-Month AA Financial Commercial Paper Rate & 6     & 2 \\
    78    & TB3MS & 3-Month Treasury Bill: & 6     & 2 \\
    79    & TB6MS & 6-Month Treasury Bill: & 6     & 2 \\
    80    & GS1   & 1-Year Treasury Rate & 6     & 2 \\
    81    & GS5   & 5-Year Treasury Rate & 6     & 2 \\
    82    & GS10  & 10-Year Treasury Rate & 6     & 2 \\
    83    & AAA   & Moodys Seasoned Aaa Corporate Bond Yield & 6     & 2 \\
    84    & BAA   & Moodys Seasoned Baa Corporate Bond Yield & 6     & 2 \\
    85    & COMPAPFFx & 3-Month Commercial Paper Minus FEDFUNDS & 6     & 1 \\
    86    & TB3SMFFM & 3-Month Treasury C Minus FEDFUNDS & 6     & 1 \\
    87    & TB6SMFFM & 6-Month Treasury C Minus FEDFUNDS & 6     & 1 \\
    88    & T1YFFM & 1-Year Treasury C Minus FEDFUNDS & 6     & 1 \\
    89    & T5YFFM & 5-Year Treasury C Minus FEDFUNDS & 6     & 1 \\
    90    & T10YFFM & 10-Year Treasury C Minus FEDFUNDS & 6     & 1 \\
    91    & AAAFFM & Moodys Aaa Corporate Bond Minus FEDFUNDS & 6     & 1 \\
    92    & BAAFFM & Moodys Baa Corporate Bond Minus FEDFUNDS & 6     & 1 \\
    93    & EXSZUSx & Switzerland / U.S. Foreign Exchange Rate & 6     & 5 \\
    94    & EXJPUSx & Japan / U.S. Foreign Exchange Rate & 6     & 5 \\
    95    & EXUSUKx & U.S. / U.K. Foreign Exchange Rate & 6     & 5 \\
    96    & EXCAUSx & Canada / U.S. Foreign Exchange Rate & 6     & 5 \\
    97    & WPSFD49207 & PPI: Finished Goods & 7     & 6 \\
    98    & WPSFD49502 & PPI: Finished Consumer Goods & 7     & 6 \\
    99    & WPSID61 & PPI: Intermediate Materials & 7     & 6 \\
    100   & WPSID62 & PPI: Crude Materials & 7     & 6 \\
    101   & OILPRICEx & Crude Oil, spliced WTI and Cushing & 7     & 6 \\
    102   & PPICMM & PPI: Metals and metal products: & 7     & 6 \\
    103   & CPIAUCSL & CPI : All Items & 7     & 6 \\
    104   & CPIAPPSL & CPI : Apparel & 7     & 6 \\
    105   & CPITRNSL & CPI : Transportation & 7     & 6 \\
    106   & CPIMEDSL & CPI : Medical Care & 7     & 6 \\
    107   & CUSR0000SAC & CPI : Commodities & 7     & 6 \\
    108   & CUSR0000SAD & CPI : Durables & 7     & 6 \\
    109   & CUSR0000SAS & CPI : Services & 7     & 6 \\
    110   & CPIULFSL & CPI : All Items Less Food & 7     & 6 \\
    111   & CUSR0000SA0L2 & CPI : All items less shelter & 7     & 6 \\
    112   & CUSR0000SA0L5 & CPI : All items less medical care & 7     & 6 \\
    113   & PCEPI & Personal Cons.  Expend.:  Chain Index & 7     & 6 \\
    114   & DDURRG3M086SBEA & Personal Cons.  Exp:  Durable goods & 7     & 6 \\
    115   & DNDGRG3M086SBEA & Personal Cons.  Exp:  Nondurable goods & 7     & 6 \\
    116   & DSERRG3M086SBEA & Personal Cons.  Exp:  Services & 7     & 6 \\
    117   & CES0600000008 & Avg Hourly Earnings :  Goods-Producing & 2     & 6 \\
    118   & CES2000000008 & Avg Hourly Earnings :  Construction & 2     & 6 \\
    119   & CES3000000008 & Avg Hourly Earnings :  Manufacturing & 2     & 6 \\
    120   & UMCSENTx & Consumer Sentiment Index & 4     & 2 \\
    121   & DTCOLNVHFNM & Consumer Motor Vehicle Loans Outstanding & 5     & 6 \\
    122   & DTCTHFNM & Total Consumer Loans and Leases Outstanding & 5     & 6 \\
    123   & INVEST & Securities in Bank Credit at All Commercial Banks & 5     & 6 \\

\end{longtable}%


\section*{Appendix B: Estimation of $k_{0}$ through Cross Validation}

The observation that the choice of the parameter $k_0$ has an influence over the performance of the model and, in particular, over the performance of the procedure for factor selection, gives us a motivation for the introduction of a criterion for the determination of an optimal value for $k_{0}$ in our empirical analysis.

As shown in the previous section, the choice of $k_{0}$ appears to have an impact on the FMMDE forecasting performance. Since no explicit rule for the determination of the parameter is provided in the original paper, we resort to a cross validation procedure, adapted for the case of time series forecasting, for determining the optimal value of $k_{0}$. 

The main idea underlying cross validation is to use a sample of past predicted values, referred to as validation set, to perform an ex-ante selection of the value $k_{0}$ to be used in subsequent forecasts. Specifically, given a discrete set of possible values for $k_{0}$, we select the parameter that minimizes the mean squared forecast error computed over the set of predictions included into the validation set. The value $k^{*}_{0}$ of choice will be then used to obtain the first forecast value outside of the validation set. The procedure is repeatedly implemented to obtain optimal forecast values at all time intervals: accordingly, a proper sequence of validation sets needs to be defined, where each of these will be employed in selecting the $k^{*}_0$ to be used in the first forecast outside of the set.

Let $t_{0,h}^{cv}$ and $t_{last,j}^{cv}$ be the time indexes corresponding to the first and last prediction included in the validation set, respectively. The latter, in particular, will be defined according to the rule $t_{last,j}^{cv}=t_{1}^{cv}+j$, for a certain $j=\left[0,1,...,577\right]$, and a pre-specified initial date $t_{1}^{cv}$, corresponding to the month of December 1975 by assumption. Let us denote as $\mathcal{I}_{h,j}^{cv}$ the validation set corresponding to the time interval $[t_{0,h}^{cv}, \text{\\ \\ \\  } t_{last,j}^{cv}]$.

Throughout the estimation $t_{0,h}^{cv}$ is assumed be fixed and determined in accordance with the selected forecast horizon: for instance, it will correspond to the date of January 1970 for $h=1$, to March 1970 for $h=3$, to June 1970 for $h=6$, and so on. At the same time, $t_{last,j}^{cv}$ will be progressively shifting according to the index $j$, so that we actually are able to consider a sequence of validation sets $\left\{\mathcal{I}_{h,j}^{cv}\right\}^{577}_{j=0} $ defined by a progressively expanding collection of forecasts. In this sense, $\mathcal{I}_{h,j+1}^{cv}$ will in fact largely coincide with $\mathcal{I}_{h,j}^{cv}$, additionally including the forecast value obtained at time $t_{last,j}^{cv}$, denoted as $\hat{y}^{h}_{t_{last,j}}$.

\medskip
As a matter of example, consider the initial validation set $\mathcal{I}_{h,0}^{cv}$, defined for the time interval $[t_{0,h}^{cv}, \text{\\ \\ \\  } t_{last,0}^{cv}]$. The first forecast value included into the set, $\hat{y}^{h}_{t_{0,h}}$, is obtained according to the methodology described in Section 4 using a fraction of the original sample ranging from January 1959 to December 1969.
Similarly, subsequent elements included into the set are obtained by following the usual recursive forecasting  procedure, so that the second element of the set, $\hat{y}^{h}_{t_{0,h}+1}$, will be obtained by employing observations ranging from January 1959 to January 1970. This goes on until the value of $\hat{y}^{h}_{t_{last,0}}$ is added to $\mathcal{I}_{h,0}^{cv}$: at this point the validation set is complete and can be used to obtain an estimate for $k_{0}$. In particular, we will select the value of $k_{0} \in \left\{1,...,25\right\}$ that minimizes the mean squared forecast error computed by considering all values in $\mathcal{I}_{h,0}^{cv}$. Such value $k^{*}_{0}$ will be then used to obtain the first forecast value outside of $\mathcal{I}_{h,0}^{cv} $, that is, the forecast value corresponding to the time index $t^{cv}_{\left(last+h\right),0}$. The described set of passages is repeated for the entire sequence of validation sets $\left\{\mathcal{I}_{h,j}^{cv}\right\}^{577}_{j=0} $.

The detailed procedure for the selection of $k_0$ is presented in Table 2. Each panel represents the time location for constructing the forecast in real time. Each row shows the dates of the Validation set, the construction of cross validation mean squared error $\left(CV-MSE\right)$ and the real time forecast obtained using the optimal value of $k_0^{*}$.

\begin{landscape}
	\begin{table}
		\centering
		\renewcommand*{\arraystretch}{2} \centering
		\scalebox{0.72}{
			\begin{tabular}{cccccc}
				\toprule
				& {\Large \textbf{$h=1$}} & {\Large \textbf{$h=3$}} & {\Large \textbf{$h=6$}} & {\Large \textbf{$h=12$}} & {\Large \textbf{$h=24$}}\\
				\cmidrule{2-6}          &       &       &       &       &  \\
				& \multicolumn{5}{c}{\Large \textbf{$December-1975$}} \\
				\cmidrule{2-6}    Validation Set & $\mathcal{I}^{cv} = \{Jan70,\, Dec75 \}$ & $\mathcal{I}^{cv} = \{Mar70,\, Dec75 \}$ & $\mathcal{I}^{cv} = \{Jun70,\, Dec75 \}$ & $\mathcal{I}^{cv} = \{Dec70,\, Dec75 \}$ & $\mathcal{I}^{cv} = \{Dec71,\, Dec75 \}$ \\
				$CV-MSE(k_0)$ & $(73-h)^{-1} \sum_{\tau = Jan70}^{Dec75} \left(y_{\tau}- \hat{y}_{\tau}\right)^{2}$ & $(73-h)^{-1} \sum_{\tau = Mar70}^{Dec75} \left(y_{\tau}- \hat{y}_{\tau}\right)^{2}$ & $(73-h)^{-1} \sum_{\tau = Jun70}^{Dec75} \left(y_{\tau}- \hat{y}_{\tau}\right)^{2}$ & $(73-h)^{-1} \sum_{\tau = Dec70}^{Dec75} \left(y_{\tau}- \hat{y}_{\tau}\right)^{2}$ & $(73-h)^{-1} \sum_{\tau = Dec71}^{Dec75} \left(y_{\tau}- \hat{y}_{\tau}\right)^{2}$ \\
				Real-Time Forecast & $\hat{y}_{Jan-1976,k_{0}^{*}}$ & $\hat{y}_{Mar-1976,k_{0}^{*}}$ & $\hat{y}_{Jun-1976,k_{0}^{*}}$ & $\hat{y}_{Dic-1976,k_{0}^{*}}$ & $\hat{y}_{Dic-1977,k_{0}^{*}}$ \\
				\midrule
				&       &       &       &       &  \\
				& \multicolumn{5}{c}{\Large \textbf{$January-1976$}} \\
				\cmidrule{2-6}    Validation Set & $\mathcal{I}^{cv} = \{Jan70,\, Jan76 \}$ & $\mathcal{I}^{cv} = \{Mar70,\, Jan76 \}$ & $\mathcal{I}^{cv} = \{Jun70,\, Jan76 \}$ & $\mathcal{I}^{cv} = \{Dec70,\, Jan76 \}$ & $\mathcal{I}^{cv} = \{Dec71,\, Jan76 \}$ \\
				$CV-MSE(k_0)$ & $(74-h)^{-1} \sum_{\tau = Jan70}^{Jan76} \left(y_{\tau}- \hat{y}_{\tau}\right)^{2}$ & $(74-h)^{-1} \sum_{\tau = Mar70}^{Jan76} \left(y_{\tau}- \hat{y}_{\tau}\right)^{2}$ & $(74-h)^{-1} \sum_{\tau = Jun70}^{Jan76} \left(y_{\tau}- \hat{y}_{\tau}\right)^{2}$ & $(74-h)^{-1} \sum_{\tau = Dec70}^{Jan76} \left(y_{\tau}- \hat{y}_{\tau}\right)^{2}$ & $(74-h)^{-1} \sum_{\tau = Dec71}^{Jan76} \left(y_{\tau}- \hat{y}_{\tau}\right)^{2}$ \\
				Real-Time Forecast & $\hat{y}_{Feb-1976,k_{0}^{*}}$ & $\hat{y}_{Apr-1976,k_{0}^{*}}$ & $\hat{y}_{Jul-1976,k_{0}^{*}}$ & $\hat{y}_{Jan-1976,k_{0}^{*}}$ & $\hat{y}_{Jan-1977,k_{0}^{*}}$ \\
				
				&   $\vdots$     &    $\vdots$    &  $\vdots$      &    $\vdots$    & $\vdots$ \\
				&   $\vdots$     &    $\vdots$    &  $\vdots$      &    $\vdots$    &  $\vdots$\\
				
				& \multicolumn{5}{c}{\Large \textbf{$December-2017$}} \\
				\cmidrule{2-6}    Validation Set & $\mathcal{I}^{cv} = \{Jan70,\, Dec2017 \}$ & $\mathcal{I}^{cv} = \{Mar70,\, Dec2017 \}$ & $\mathcal{I}^{cv} = \{Jun70,\, Dec2017 \}$ & $\mathcal{I}^{cv} = \{Dec70,\, Dec2017\}$ & $\mathcal{I}^{cv} = \{Dec71,\, Dec2017 \}$ \\
				$CV-MSE(k_0)$ & $(577-h)^{-1} \sum_{\tau = Jan70}^{Dec2017} \left(y_{\tau}- \hat{y}_{\tau}\right)^{2}$ & $(577-h)^{-1} \sum_{\tau = Mar70}^{Dec2017} \left(y_{\tau}- \hat{y}_{\tau}\right)^{2}$ & $(577-h)^{-1} \sum_{\tau = Jun70}^{Dec2017} \left(y_{\tau}- \hat{y}_{\tau}\right)^{2}$ & $(577-h)^{-1} \sum_{\tau = Dec70}^{Dec2017} \left(y_{\tau}- \hat{y}_{\tau}\right)^{2}$ & $(577-h)^{-1} \sum_{\tau = Dec71}^{Dec2017} \left(y_{\tau}- \hat{y}_{\tau}\right)^{2}$ \\
				Real-Time Forecast & $\hat{y}_{Jan-2018,k_{0}^{*}}$ & $\hat{y}_{Mar-2018,k_{0}^{*}}$ & $\hat{y}_{Jun-2018,k_{0}^{*}}$ & $\hat{y}_{Dec-2018,k_{0}^{*}}$ & $\hat{y}_{Dec-2019,k_{0}^{*}}$ \\
				\bottomrule
		\end{tabular}}
		\label{tab:CvTab}%
		\caption{Detailed proceudre for the selection of $k_{0}$ through cross validation}
	\end{table}%
\end{landscape}